\documentclass[reqno,12pt]{article}
\usepackage{amsthm,amsmath,amssymb,graphicx,setspace,harvard,verbatim,subfigure}
\usepackage[small]{caption}
\usepackage{wrapfig}
\usepackage{multirow}

\renewcommand{\arraystretch}{1}

\newcommand{\bzero}{\mbox{\boldmath $0$}}

\newcommand{\bK}{\mbox{\boldmath $K$}}
\newcommand{\bm}{\mbox{\boldmath $m$}}

\newcommand{\bt}{\mbox{\boldmath $t$}}

\newcommand{\bw}{\mbox{\boldmath $w$}}
\newcommand{\bx}{\mbox{\boldmath $x$}}

\newcommand{\by}{\mbox{\boldmath $y$}}
\newcommand{\bz}{\mbox{\boldmath $z$}}
\newcommand{\bA}{\mbox{\boldmath $A$}}

\newcommand{\bI}{\mbox{\boldmath $I$}}

\newcommand{\bP}{\mbox{\boldmath $P$}}

\newcommand{\bV}{\mbox{\boldmath $V$}}

\newcommand{\cB}{{\cal B}}
\newcommand{\cC}{{\cal C}}

\newcommand{\cE}{{\cal E}}

\newcommand{\cL}{{\cal L}}

\newcommand{\cN}{{\cal N}}

\newcommand{\cY}{{\cal Y}}
\newcommand{\cZ}{{\cal Z}}

\newcommand{\eps}{\varepsilon}

\newcommand{\bone}{\mbox{\boldmath $1$}}
\newcommand{\beps}{\mbox{\boldmath $\varepsilon$}}

\newcommand{\bbeta}{\mbox{\boldmath $\beta$}}

\newcommand{\btheta}{\mbox{\boldmath $\theta$}}

\newcommand{\bxi}{\mbox{\boldmath $\xi$}}

\newcommand{\bgamma}{\mbox{\boldmath $\gamma$}}
\newcommand{\bdelta}{\mbox{\boldmath $\delta$}}

\newcommand{\bpi}{\mbox{\boldmath $\pi$}}

\newcommand{\bSigma}{\mbox{\boldmath $\Sigma$}}

\newcommand{\bLambda}{\mbox{\boldmath $\Lambda$}}
\newcommand{\bPsi}{\mbox{\boldmath $\Psi$}}
\newcommand{\bPhi}{\mbox{\boldmath $\Phi$}}
\newcommand{\bOmega}{\mbox{\boldmath $\Omega$}}

\newcommand{\bUpsilon}{\mbox{\boldmath $\Upsilon$}}

\newcommand{\E}{\mbox{E}}

\newcommand{\Cor}{\mbox{Cor}}
\newcommand{\Wishart}{\mbox{Wishart}}
\newcommand{\miss}{\mbox{\scriptsize miss}}
\newcommand{\obs}{\mbox{\scriptsize obs}}
\newcommand{\logit}{\mbox{logit}}

\newcommand{\bdm}{\begin{displaymath}}
\newcommand{\edm}{\end{displaymath}}
\newcommand{\beq}{\begin{equation}}
\newcommand{\eeq}{\end{equation}}

\long\def\symbolfootnote[#1]#2{\begingroup%
\def\thefootnote{\fnsymbol{footnote}}\footnote[#1]{#2}\endgroup}

\begin{document}

\pagenumbering{arabic}
\begin{center}
{\singlespacing
\begin{Large}{\bf
 Calibration of Computational Models with Categorical Parameters and Correlated Outputs via Bayesian Smoothing Spline ANOVA.\\} 
\vspace{.25in}
\end{Large}

Curtis B. Storlie$^\dag$, William A. Lane$^\ddag$, Emily M. Ryan$^\ddag$, James R. Gattiker$^\dag$, \\David M. Higdon$^\dag$ \\[.15in]

$^\dag$ Los Alamos National Laboratory \hspace{.2in} $^\ddag$ Boston University\\

\begin{abstract}

It has become commonplace to use complex computer models to predict outcomes in regions where data does not exist.  Typically these models need to be calibrated and validated using some experimental data, which often consists of multiple correlated outcomes.  In addition, some of the model parameters may be categorical in nature, such as a pointer variable to alternate models (or submodels) for some of the physics of the system.
Here we present a general approach for calibration in such situations where an emulator of the computationally demanding models and a discrepancy term from the model to reality are represented within a Bayesian Smoothing Spline (BSS) ANOVA framework.  The BSS-ANOVA framework has several advantages over the traditional Gaussian Process, including ease of handling categorical inputs and correlated outputs, and improved computational efficiency.  Finally this framework is then applied to the problem that motivated its design; a calibration of a computational fluid dynamics model of a bubbling fluidized which is used as an absorber in a CO$_2$ capture system.

\vspace{.075in}

\vspace{.075in}
\noindent
{\em Running title}: BSS-ANOVA Model Calibration

\vspace{.075in}
\noindent
{\em Corresponding Author}: Curtis Storlie, \verb1storlie@lanl.gov1

\end{abstract}
}

\end{center}


\vspace{-.35in}
\section{Introduction}
\vspace{-.1in}

The analysis of many physical and engineering problems (e.g., climate change, nuclear reactor performance, fluid transport, and carbon capture systems) involves running complex computational models (i.e., simulators).  The importance of uncertainty quantification (UQ) as an integral component to the overall analysis of complex computer models is almost universally recognized \cite{Beck87,Helton00,Kennedy01,Oakley04,Helton06,Helton07,Storlie09b}.  Model calibration and model assessment (i.e., validation) are two important components of UQ that are addressed in this paper in the context of computational fluid dynamics (CFD) models as part of the Department of Energy's (DOE's) Carbon Capture Simulation Initiative (CCSI).


In particular we investigate a bubbling fluidized bed with immersed horizontal heat transfer tubes, based off the experimental work of \citeasnoun{Kim03}.  Fluidized beds are widely used in chemical engineering systems and processes (e.g., combustion, mixing, polymerization, and carbon capture) 
\cite{Asegehegn11}.  This particular problem is a ``unit'' problem within a larger set of problems being investigated through CCSI.  The main goal of this work was to develop tools and techniques for UQ appropriate for use with CFD models.  Eventually, the general UQ methodology will be used for further calibration on laboratory and pilot scale systems, and finally for uncertainty propagation to a full scale CO$_2$ capture system.  

\begin{figure}[t!]
\begin{center}
\caption{(a) Bubbling Fluidized Bed Experimental Setup and (b) MFIX simulation setup.}
\vspace{-.05in}
\includegraphics[width=.90\textwidth]{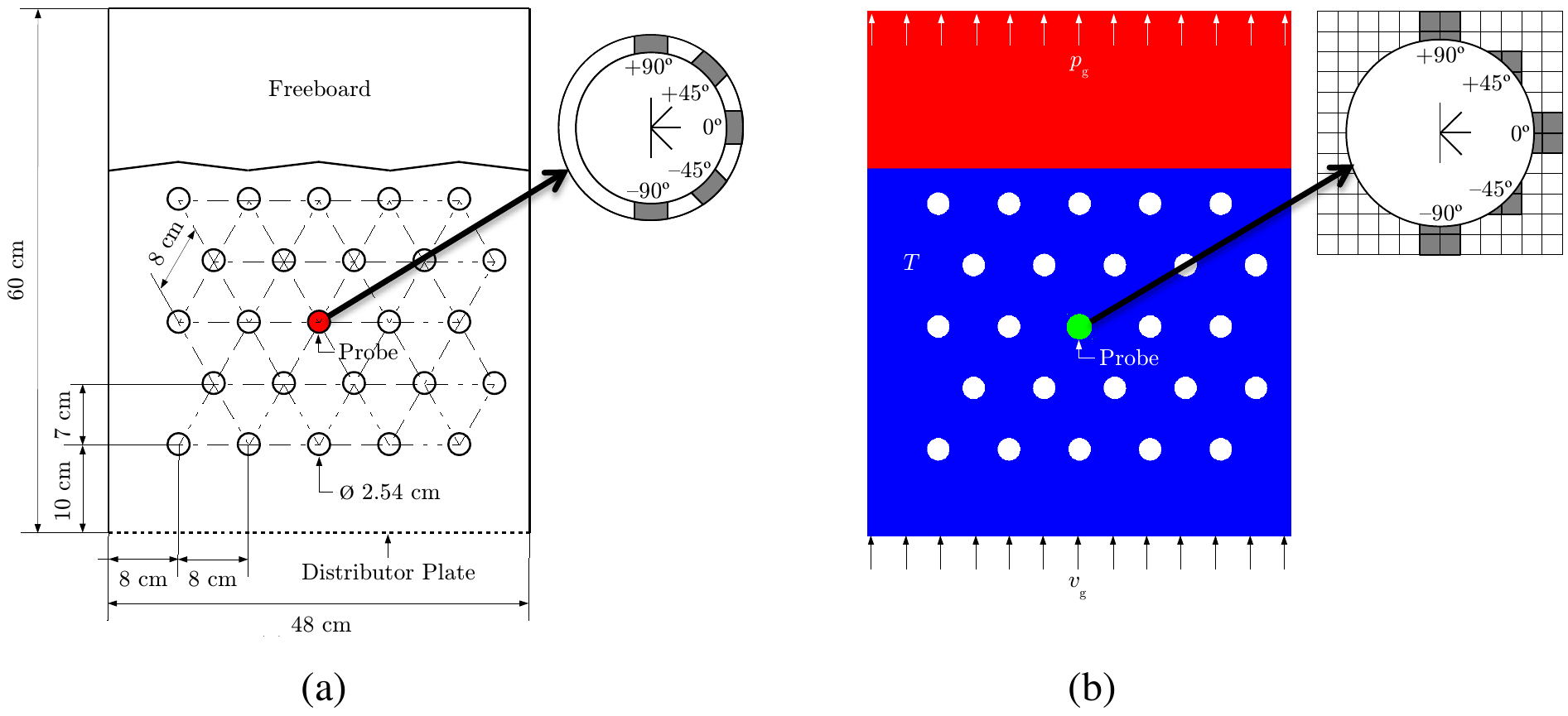}
\label{fig:bub_bed}
\end{center}
\vspace{-.2in}
\end{figure}

Below we provide an overview of the experimental setup and the CFD simulations.  Full details of the problem setup and the variables involved can be found in  \citeasnoun{Lane13}. The experimental setup of \citeasnoun{Kim03} measured $48 \times 60 \times 34$ cm (width $\times$ height $\times$ depth). A staggered tube bundle of twenty-five 2.54 cm diameter tubes was fixed 10 cm above the base (measured to the center of the bottom row of tubes). A full schematic of the setup is shown in Figure~\ref{fig:bub_bed}. The central tube (probe) was fitted with optical and thermocouple sensors at angular locations of {-90, -45, 0, 45, 90}$^\circ$, as seen in Figure~\ref{fig:bub_bed}. Air was pumped through a distributor plate located on the bottom causing bubbles to form and fluidization to occur. As the bubbles moved upward through the bed, their behavior was recorded by the probe and used to calculate bubble frequency, bubble phase fraction, and contacting time.

The open source CFD code Multiphase Flow with Interphase eXchanges (MFIX) \cite{Benyahia12},  was used to simulate the fluidized bed.  
A two-dimensional approximation was used to generate the CFD simulation domain.  A planar domain measuring $48 \times 60$ cm (width $\times$ height) was used to approximate the middle of the system (depth = 17 cm). The bottom and top boundaries were set as constants: velocity inflow and pressure outflow, respectively. The left and right boundaries were treated as no-slip-walls (zero velocity, zero pressure gradient). Physical properties of the system (e.g., particle diameter, particle and gas densities, etc.) were set to the experimental values reported. Simulations were run were for 60 sec real-time, of which the first 30 sec was ignored to avoid transient start-up effects.  The experimental (and simulator) outputs, inputs, and simulation model parameters are summarized in Table~\ref{tab:data_descr}.  The first five elements of the model parameter vector $\bt$ are continuous, while the last element is a discrete parameter with three levels. Further details about the experimental and simulation setups, including discussion of model parameters and post-processing simulator results, etc., can be found in \citeasnoun{Kim03} and \citeasnoun{Lane13}, respectively.

\begin{table}[t!]
\caption{Summary of Inputs, Outputs, and CFD model parameters (along with prior distributions and abbreviations used in Figures~\ref{fig:SA}~and~\ref{fig:posterior_theta}) for the bubbling fluidized bed calibration.}
\vspace{-.1in}
\framebox{
\small
\begin{minipage}{1\textwidth}
\begin{itemize}
 \setlength{\itemindent}{-.275in}
 \setlength{\leftmargin}{-.275in}
\item[] $\!\!\!\!\!\!\!$Outputs $\by$:\\[-.33in]
 \begin{itemize}
 \setlength{\itemindent}{-.3in}
 \setlength{\leftmargin}{-.3in}
 \item[$y_1$]: Bubble Frequency $-$ measured in hertz, experimentally observed at angles $\{-90, -45, 0, 45, 90 \}^\circ$ and velocities $\{5.5, 7.0, 11.0, 12.6\}$ cm/sec.\\[-.25in]
 \item[$y_2$]: Bubble Phase Fraction $-$ the proportion of time a bubble is present, experimentally observed at angles $\{-90, -45, 0, 45, 90 \}^\circ$ and velocity $12.6$ cm/sec.\\[-.275in]
 \end{itemize}
\item[] $\!\!\!\!\!\!\!$Inputs $\bx$:\\[-.33in]
 \begin{itemize}
 \setlength{\itemindent}{-.3in}
 \setlength{\leftmargin}{-.3in}
 \item[$x_1$]: Gas Velocity, $[5.5, 16.1]$ cm/sec.\\[-.25in]
 \item[$x_2$]: Angular Location on Tube, $[-90, 90]^\circ$\\[-.25in]
 \end{itemize}
\item[] $\!\!\!\!\!\!\!$CFD Model Parameters $\bt$:\\[-.33in]
 \begin{itemize}
 \setlength{\itemindent}{-.3in}
 \setlength{\leftmargin}{-.3in}
 \item[$t_1$]: Coefficient of restitution, particle-particle ({\em Res.PP}), $[0.8,1.0]$\\[-.225in]
 \item[$t_2$]: Coefficient of restitution, particle-wall ({\em Res.PW}), $[0.8,1.0]$\\[-.225in]
 \item[$t_3$]: Friction angle, particle-particle ({\em FricAng.PP}), $[25,45]$\\[-.225in]
 \item[$t_4$]: Friction angle, particle-wall ({\em FricAng.PW}), $[25,45]$\\[-.225in]
 \item[$t_5$]: Packed bed void fraction ({\em PBVF}), $[0.3,0.4]$\\[-.225in]
 \item[$t_6$]: Drag model$\:$({\em DragMod}), $\{$''Syamlal-O'Brien'', $\!\!$''Wen-Yu'', $\!\!$''Gidaspow''$\!\}\!\!$
 \end{itemize}
\end{itemize}
\end{minipage}
}
\label{tab:data_descr}
\vspace{-.055in}
\end{table}

Model calibration/assessment are important in problems such as this where there are experimental (or field) data from the true physical system, collected in some region of the input space (e.g., at certain flow rates and angular locations).  Often there are multiple outcomes (e.g., bubble frequency and phase fraction) of interest, and the field data measures at least some of these outcomes.  The goal is then to calibrate (i.e., find a plausible set of values for the model parameters) and validate (i.e., demonstrate acceptable predictive performance) the model, and then use the model to make predictions about outcomes in regions of the input space where data does not exist.

This goal is routinely complicated by several factors: (i) the computational models are very expensive to run, (ii) the multiple outputs of interest are correlated, (iii) there may be categorical parameters (e.g., a pointer to alternate models for some of the physics of the system), and (iv) the experimental data may contain an incomplete record for some of the outcomes. In this paper we present a general approach for computer model calibration in settings that have all of the above challenges and apply this approach to the bubbling fluidized bed problem discussed above.

For a single output of interest,  \citeasnoun{Kennedy01} and \citeasnoun{Higdon04}  provide a statistically rigorous model based framework for computer model calibration which deals with complication (i) via the use of an emulator (a statistical approximation to the computer model).  These approaches provide a posterior distribution for the model parameters and a model discrepancy term, and thus easily account for uncertainty in model predictions.

Some work has been done on emulators for multivariate output.  One class of this work considers the common problem where multidimensional model output is functional in nature, i.e., coming from locations on a Cartesian grid \cite{Kennedy01,Rougier08,Higdon08,Bayarri07}.   However, what is meant by multivariate output here is the case where simulators produce different types of output quantities simultaneously (i.e., bubble frequency and bubble phase fraction).  \citeasnoun{Conti10} tackle this problem with a multivariate GP, where the correlation structure between outputs is separable with the correlation across input space.   \citeasnoun{Hankin12} and \citeasnoun{Fricker13} extend this concept to nonseparable structures so that the input correlation lengths can be different for each output.  Our approach has some similarity, where a separable structure is assumed for each of the many functional components of the functional ANOVA decomposition of the emulator, resulting in a nonseparable covariance structure for the overall multivariate GP.
The \citeasnoun{Higdon08} approach tackles the multivariate output problem via functional principle component decomposition aimed at the first case of functional output, but can be used for multiple type output as well.  This approach was also developed within the calibration framework, and is thus the closest existing approach to what is needed here.

There has also been recent interest in the treatment of categorical parameters in computer models \cite{Qian08,Zhou11,Storlie12b}.  However, all of this work has been aimed only at developing emulators for the computer model for the purpose of sensitivity analysis (SA) or uncertainty analysis (UA).  To the best of our knowledge, there is not a calibration approach available that appropriately handles categorical model parameters.

Here we present a general approach for calibration in situations where there are multiple types of outcomes of interest observed across input space, the computer model(s) are expensive to run, there are categorical parameters to calibrate, and there are missing data.  Many multiple (competing) model situations can also be handled with this approach by treating the distinct models as levels of a categorical parameter.  An emulator of the computationally demanding model(s) and a discrepancy term from the model to reality are represented within a Bayesian Smoothing Spline (BSS) ANOVA framework.  The BSS-ANOVA framework has several advantages over the traditional Gaussian Process, including flexible handling categorical parameters, correlated outputs, and {\em linear} computational complexity in the number of simulator observations.

This calibration approach is then applied to the bubbling fluidized bed problem that motivated its design.  While some of the scientific results of this specific problem have been presented in \citeasnoun{Lane13}, the focus here is on the presentation of the statistical methodology and analysis.  Finally, the proposed approach is tested in a controlled simulation study and performance is compared to the Gaussian Process Models for Simulation Analysis (GPMSA) approach of \cite{Higdon08} which has also been extended here to allow for categorical calibration parameters.

The rest of the paper is laid out as follows.  Section~\ref{sec:method} describes the BSS-ANOVA calibration approach, while Section~\ref{sec:estimation} provides estimation details.  A detailed calibration analysis of a CFD model of a bubbling fluidized bed is provided in Section~\ref{sec:analysis}.  Section~\ref{sec:simulation} presents a simulation study comparing the proposed approach to GPMSA and  Section~\ref{sec:conclusions} concludes the paper.  This paper also has online supplementary material containing BSS-ANOVA basis representation and Markov Chain Monte Carlo (MCMC) details.

\vspace{-.2in}
\section{The BSS-ANOVA Calibration Framework}
\vspace{-.1in}
\label{sec:method}

Here a computer model calibration framework is presented where the output of interest $\by$ is a vector of multiple outcomes and some of the model parameters are categorical.  This is done with the BSS-ANOVA model \cite{Reich09} because of two favorable properties: (i) the BSS-ANOVA Gaussian Process (GP)  allows for a simple, yet flexible treatment of categorical parameters \cite{Storlie12b}, and (ii) BSS-ANOVA can be conveniently represented as a functional form leading to a computational procedure that scales {\em linearly} with the number of data points.  In comparison, the traditional squared exponential GP results in a procedure that is cubic in the number of data points.  
We begin by reviewing the traditional calibration approach, and then present the BSS-ANOVA data model for both categorical parameters and multiple outputs

\vspace{-.15in}
\subsection{Review of the Computer Model Calibration Framework}
\vspace{-.05in}
\label{sec:cal_review}
The goal of calibration is to find a plausible set of model parameter values that best reproduce the reality of experimental (or field) data.  In the traditional computer model calibration (i.e., inverse problem) setup \cite{Kennedy01} an output, $y$, from the physical system $\zeta$ is observed (with observational error) at several ($N$) locations of a “controllable” vector of inputs $\bx=[x_1,\dots,x_P]$, i.e., $y_n=\zeta(\bx_n)+\eps_n$,
$n =1,\dots,N$, where $\eps_n$ is the observation error for the $n$-th observation.  Physical reality $\zeta(\bx)$ can be approximated by a simulator (i.e., a computer model), $\eta(\bx,\bt)$ , where $\bt=[t_1,\dots,t_Q]$ is a vector of model parameters. If
fixed at an appropriate (unknown) value of $\bt=\btheta$, then  $\eta(\bx,\btheta)$ will approximate $\zeta(\bx)$. There
can also be a model form discrepancy function that admits the possibility of model bias. Therefore the model for the experimental data is
\vspace{-.2in}\beq
y_n = \eta(\bx_n, \btheta) + \delta(\bx_n)  + \eps_n,
\label{eq:data_model}
\vspace{-.2in}\eeq
$n=1,\dots,N$.  The goal is to estimate $\btheta$ and the discrepancy function $\delta$. This is typically done within a Bayesian framework \cite{Higdon04}, where a prior distribution is placed on $\btheta$ and $\delta$ and then updated by conditioning on the experimental data.

Computation of $\eta(\bx,\bt)$ is often expensive (e.g., a day or more), which makes typical Bayesian computation to obtain posterior samples for the above problem impossible. In such cases, the simulator $\eta$ is further modeled with an emulator \cite{Kennedy01,Higdon04,Reich09}. An emulator is more than just a fast surrogate for the simulator; it is a probabilistic representation of the simulator function, i.e., a stochastic process over $(\bx,\bt)$. This is also the same way the discrepancy function is modeled (over just $\bx$). Typically a Gaussian process is assumed for both $\eta$ and $\delta$ since it results in a flexible yet tractable procedure. The simulator must then be run (outside of the calibration routine) at several ($M$) design locations $(\bx^*_1,\bt^*_1),\dots,(\bx^*_{m},\bt^*_{M})$ and the resulting outputs at these points are treated as additional data. In this setting $\btheta$, $\delta$, and $\eta$ must be estimated. Thus, a non-parametric regression problem is nested inside of the inverse problem to identify $\btheta$. The emulator evaluations must match (aside from numerical error) the simulator evaluations, but it also admits uncertainty about what the simulator would output at input values far from the observed evaluations.

\vspace{-.15in}
\subsection{The BSS-ANOVA Model}
\vspace{-.05in}

Here the BSS-ANOVA framework which provides the underpinning for the proposed calibration approach is reviewed.
BSS-ANOVA is a GP with a special covariance function that explicitly models the functional components of a functional ANOVA decomposition \cite{Gu02}.  Below we provide only the details of the BSS-ANOVA model necessary to present the calibration method, while the specific details are provided in the Supplementary Material.  The BSS-ANOVA model is described below in the context of the emulator for the simulator output $\eta$, but the same model is then assumed for $\delta$ as well.

For notational convenience, denote the combination of inputs and model parameters as the vector $\bw=(\bx,\bt)$ of dimension $R=P+Q$.  The simulator is represented as
\vspace{-.15in}\beq
\eta(\bw) = \alpha_0 +\sum_{r=1}^R \eta_r(w_r) + \sum_{r<r'}^R \eta_{r,r'}(w_r, w_{r'}) + \cdots
\label{eq:eta}
\vspace{-.15in}\eeq
It is assumed that $\alpha_0 \sim N(0,\varsigma_0^2)$, and that each main effect functional component is $\eta_r \sim GP(0,\varsigma^2_r K_1)$, for some variance parameters $\varsigma^2_r$, $r=0,\dots,R$, and $K_1$ is the BSS-ANOVA covariance function described in \cite{Reich09}.  That is,
\vspace{-.2in}\beq
K_1(u,u') = B_1(u)B_1(u') + B_2(u)B_2(u') - \frac{1}{24}B_{4}(|u-u'|),
\label{eq:BSS_cov}
\vspace{-.2in}\eeq
where $B_l$ is the $l$-th Bernoulli polynomial. The covariance function in (\ref{eq:BSS_cov}) operates on the domain $[0,1]$.  Therefore inputs and parameters must be transformed to $[0,1]$ prior to analysis.
Two-way interaction functions are assumed to be $\eta_{r,r'} \sim GP(0,\varsigma^2_{r,r'} K_2)$, where 
\vspace{-.2in}
\beq
K_2((u,v), (u',v')) = K_1(u,u') K_1(v,v').
\label{eq:BSS_2way_cov}
\vspace{-.2in}
\eeq
Three-way or higher order interaction functional components can be defined similarly.  Under this construction, the resulting component GPs are such that they will satisfy the functional ANOVA constraints, e.g.,  $\int \eta_r(u)du=0$ and $\int \eta_{r,r'}(u,v)du=0$, almost surely.  Any realization from this GP also lies in first order Sobolev space, i.e., absolutely continuous with derivative in $L_2$.

\begin{wrapfigure}{r}{.455\textwidth}
\vspace{-.45in}
\begin{center}
\caption{First nine eigenfunctions from the Karhunen-Lo\'eve expansion for a main effect function from the BSS-ANOVA covariance.}
\vspace{-.05in}
\includegraphics[width=0.435\textwidth]{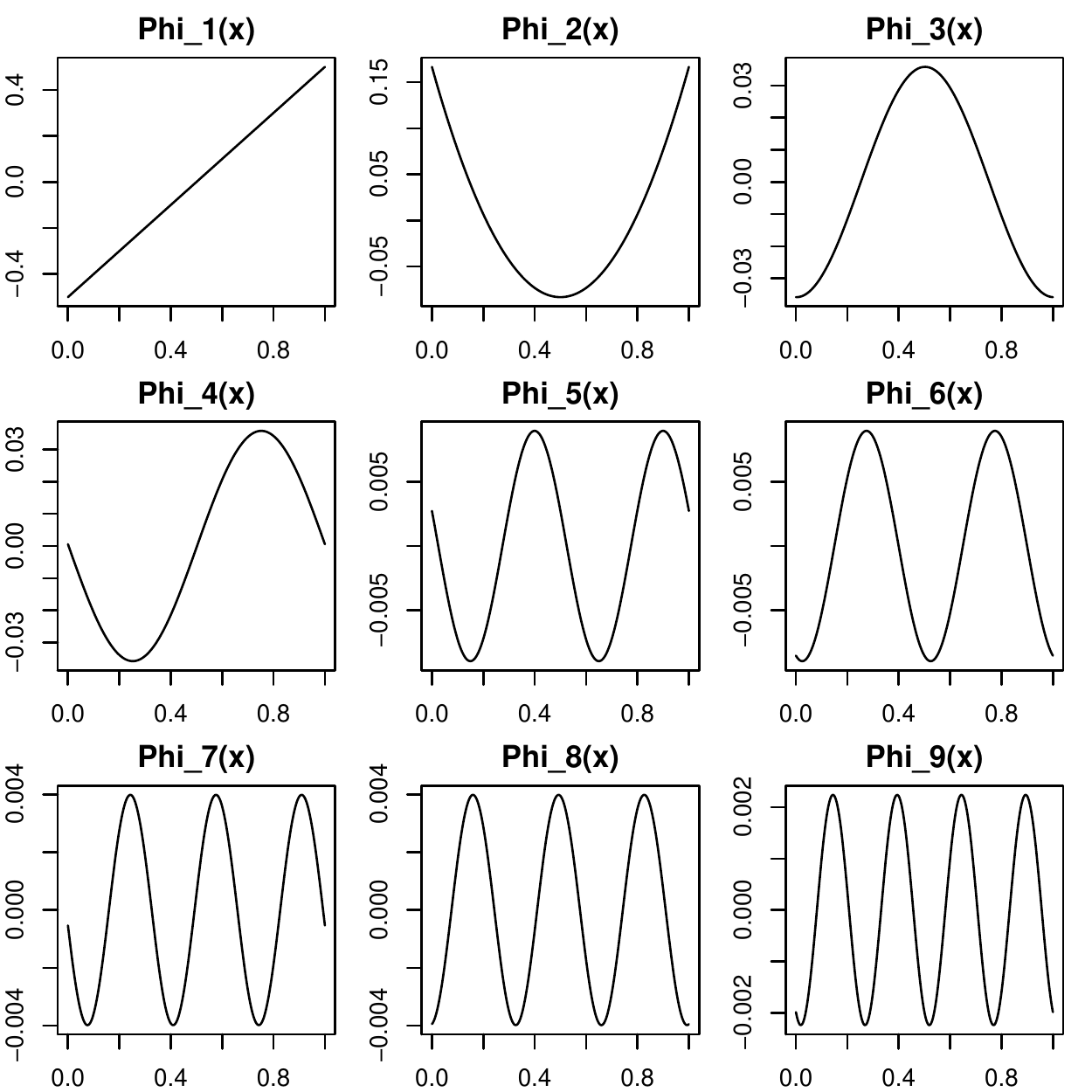} 
\vspace{-0.0in}
\label{fig:KL_basis}
\end{center}
\vspace{.7in}
\end{wrapfigure}

It was further demonstrated in \citeasnoun{Storlie13onion} that each functional component in (\ref{eq:eta}) can be further written as an orthogonal basis expansion, e.g., 
\vspace{-.1in}\beq
\eta_r(w_r) = \sum_{l=1}^\infty \alpha_{r,l} \phi_l(w_r),
\label{eq:eta_comp}
\vspace{-.1in}\eeq
where
\vspace{-.05in}\beq
\alpha_{r,l} \stackrel{iid}{\sim} {\cal N}(0,\varsigma_r^2)
\label{eq:alpha_prior}
\vspace{-.1in}\eeq
The $\phi_l$ terms in the expansion are just the eigenfunctions (scaled by the eigenvalues) in the Karhunen-Lo\'eve (KL) expansion (\citeasnoun{Berlinet04}, pp.~65-70).  The $\phi_l$ get increasingly higher frequency and have decreasingly less magnitude as depicted in Figure~\ref{fig:KL_basis}, so the expansion in (\ref{eq:eta_comp}) can be truncated at some value $L$.  The choice of $L$ is not critical, as the model will be practically identical for different $L$ provided it is large enough.  In our experience, results have been only negligibly different for $L \geq 25$. 

The same decomposition in (\ref{eq:eta_comp}) and (\ref{eq:alpha_prior}) can be used for two-way and higher interactions as well.  In fact, the $\phi_l$ for two way interactions are simply pairwise products of the corresponding main effect basis functions and similarly for three way and higher interactions.  
In many problems it is sufficient to include only main effects and two-way interactions.  In our experience with this approach, allowing three-way interactions between two elements of the input vector $\bx$ with one element of model parameter vector $\theta$ provides an entirely sufficient model fit in most cases.  

Regardless of the specific functional components chosen for inclusion, the overall model in (\ref{eq:eta}) can be written in general as 
\vspace{-.15in}\beq
\eta(\bw) =  \sum_{j=1}^J \sum_{l=1}^{L^\eta_j} \beta_{j,l} \varphi_{j,l}(\bw),
\label{eq:eta2}
\vspace{-.2in}\eeq
where (i) $\beta_{j,l} \stackrel{ind}{\sim} {\cal N}(0,\lambda_j^2)$, (ii) $j$ indexes over the $J$ functional components included in the emulator, and (iii) $l$ indexes over the number of basis functions $L^\eta_j$ used for the $j$-th functional component of the emulator.  The $\beta_{j,l}$, $\varphi_{j,l}$, and $\lambda_j$ would correspond to a particular $\alpha_{r,l}$, $\phi_l(w_r)$, $\varsigma_r^2$ in (\ref{eq:eta_comp}), respectively, depending on the functional component represented by $j$.  More specific details of the decomposition of the BSS-ANOVA GP into the linear model in (\ref{eq:eta2}) are provided in the Supplementary Material.  

In a completely analogous fashion, the discrepancy function can be written as
\vspace{-.15in}\beq
\delta(\bx) =  \sum_{k=1}^K \sum_{l=1}^{L^\delta_k} \gamma_{k,l} \psi_{k,l}(\bx),
\label{eq:delta2}
\vspace{-.2in}\eeq
where (i) $\gamma_{k,l} \stackrel{ind}{\sim} {\cal N}(0,\omega_k^2)$, (ii) $k$ indexes over the $K$ functional components included in the discrepancy, and (iii) $l$ indexes over the number of basis functions $L^\delta_k$. 

The reason that the KL decomposition in (\ref{eq:eta2}) and (\ref{eq:delta2}) is very beneficial here is that the basis functions for each functional component (e.g., $\varphi_{j,l}$) do not change with the covariance function parameters ($\lambda_j$). Thus, the KL decomposition of $K_1$ needs to be done one time on a dense grid in one dimension as described in the Supplementary Material, and there is {\bf no} matrix decomposition required during the MCMC estimation procedure.  Thus, for estimation purposes, the models for $\eta$ and $\delta$ in (\ref{eq:eta2}) and (\ref{eq:delta2}) are just linear models, resulting in an $O(M+N)$ computational procedure (where recall $M$ is the number of simulator runs, and $N$ is the number of experimental observations).  Computational details are discussed further in Section~\ref{sec:estimation} and the Supplementary Material.

%

\vspace{-.15in}
\subsection{Categorical Parameters}
\vspace{-.05in}
\label{sec:cat_params}

The BSS-ANOVA framework is also amenable to unordered categorical predictors as implemented in \cite{Storlie12b}.  Assume $w_r \in \{1,2,...,G_r\}$ is categorical and so the main effect is defined by $G_r$ discrete values $\rho_{r,u} = \eta_r(u)$.  The BSS-ANOVA model for a categorical input/parameter reduces to the familiar discrete level ANOVA model where $\rho_{r,u}$ is the effect for factor level $u$.  A Bayesian ANOVA model would need to specify a prior for the $\rho_{r,u}$.  To keep with the functional ANOVA construction (and identify the intercept), the sum-to-zero constraint $\sum_{u=1}^{G_r}\rho_{r,u}=0$ must be enforced.  
This is accomplished by specifying a Gaussian distribution for the $\rho_{r,u}$ conditional on $\sum_{u=1}^{G_r}\rho_{r,u}=0$.  
This results in the singular, mean-zero Gaussian process $\eta_r(w_r) \sim GP(0,\lambda_r^2 K_d)$ over the discrete domain $\{1,2,...,G_r\}$, where
\vspace{-.2in}\beq
K_d(u,u') = \frac{G_r-1}{G_r}I(u=u') -\frac{1}{G_r}I(u\ne u'),
\label{eq:cov_cat}
\vspace{-.1in}\eeq
and $I(\cdot)$ is the indicator function.

Such a model can be equivalently represented with the basis expansion
\vspace{-.2in}\beq
\eta_r(w_r) = \sum_{l=1}^{G_r}\alpha_{r,l} \left[\frac{G_r-1}{G_r}I(w_r=l) - \frac{1}{G_r}I(w_r \neq l) \right]
\label{eq:cat_comp}
\vspace{-.1in}\eeq
where
\vspace{-.1in}\beq
\alpha_{r,l} \stackrel{iid}{\sim} {\cal N}(0,\lambda_r^2)
\label{eq:alpa_prior_cat}
\vspace{-.1in}\eeq
This representation in a purely additive BSS-ANOVA model would imply just a simple vertical shift when changing the value of the categorical parameter $w_r$.  Interactions including categorical predictors with the covariance given in (\ref{eq:cov_cat}) are handled no differently than interactions between continuous predictors, i.e., by taking products of the main effect covariance functions, or equivalently by taking pairwise products of the respective basis functions from each covariance function.  


In this model, outputs can behave differently across categories (to the extent that the simulator runs suggest this) but it is still encouraged to be similar. This is because the

\begin{wrapfigure}{r}{.455\textwidth}
\vspace{-.05in}
\centering
\caption{BSS-ANOVA emulator fit (lines) along with simulator output (points). Bubble Frequency is plotted against angular location for each of three different drag models (averaged across the other 5 model parameters. The models are similar for the different drag models and this is being leveraged in the estimation.}
\vspace{-.1in}
\includegraphics[width=.425\textwidth]{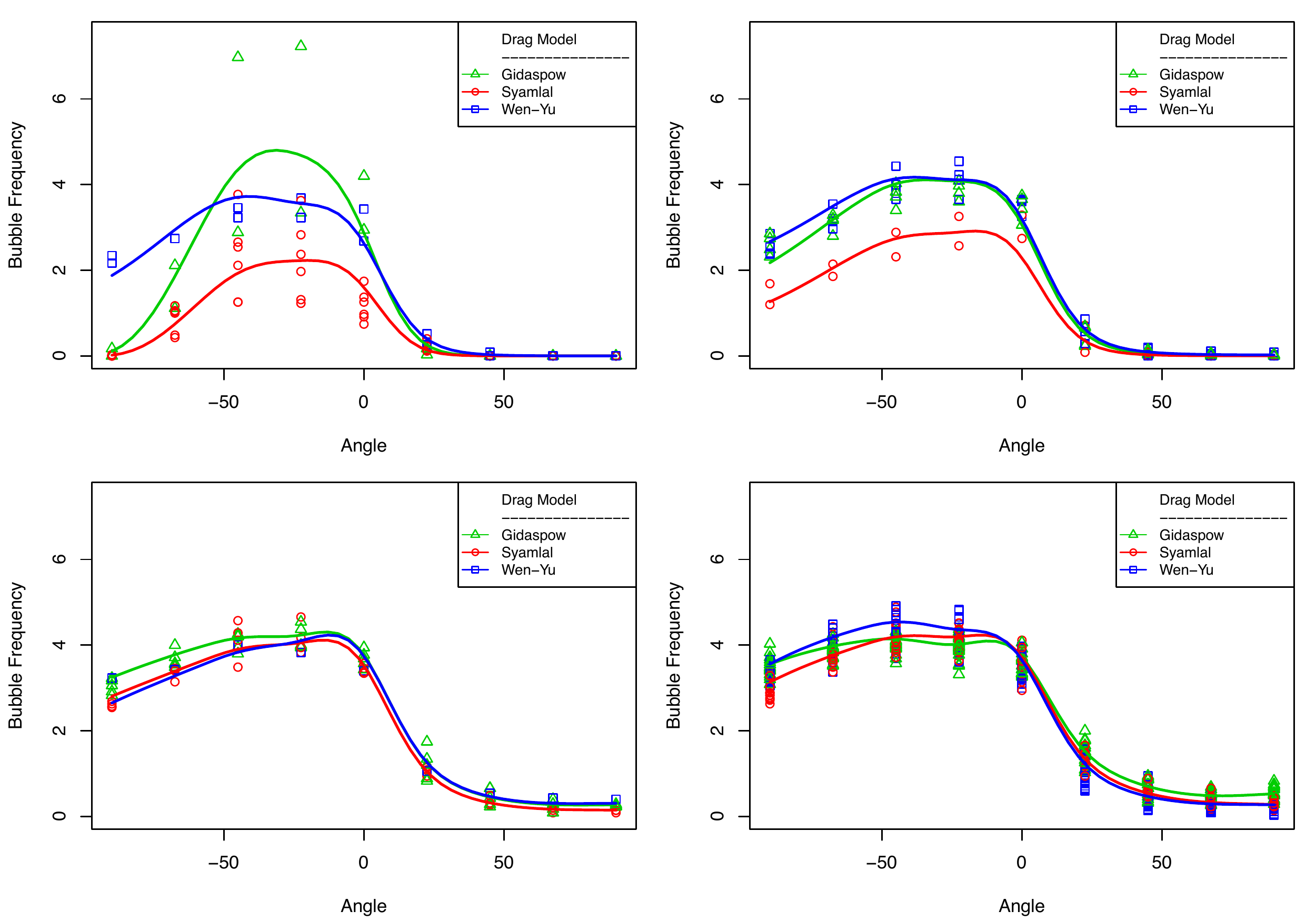}
\label{fig:cat_emulator}
\vspace{-.2in}
\end{wrapfigure}

\noindent
normal, mean zero prior on the $\alpha_{r,l}$ will have the effect of penalizing a deviation from the model in which all categories of $w_r$ produce the same output. Since they are encouraged to be similar, each category borrows strength across other categories to estimate the output function within that category. An example of this behavior can be seen in Figure~\ref{fig:cat_emulator}.  Specific details about basis construction for continuous and categorical parameters (or inputs) are provided in the Supplementary Material.


\vspace{-.15in}
\subsection{Multiple Correlated Outputs}
\vspace{-.05in}

A multivariate output of the physical system $\by=[y_1,\dots,y_C]^T$ 
is modeled as
\vspace{-.2in}\beq
\by_n = \eta(\bx_n, \btheta) + \delta(\bx_n)  + \beps_n,
\label{eq:MV_data_model}
\vspace{-.2in}\eeq
where $\eta(\bx_n, \btheta) = [\eta_1(\bx_n, \btheta), \dots,\eta_C(\bx_n, \btheta)]^T$and similarly for $\delta(\bx_n)$.  Thus, the emulator for $\eta$, the discrepancy $\delta$, and the measurement errors $\beps_n$ need a multivariate representation to appropriately account for correlation among the multiple outputs.

Assume the BSS-ANOVA basis function representation for the $j$-th functional component of the $c$-th output as in (\ref{eq:eta2}), i.e.,  
\vspace{-.15in}\beq
\eta_{j,c}(\bw_j) = \sum_{l=1}^L \beta_{j,l,c} \varphi_{j,l}(\bw_j), \;\; c=1,\dots,C,
\label{eq:MV_eta}
\vspace{-.15in}\eeq
where the basis functions $\varphi_{j,l}$ are the same for each output, and the $\beta_{j,l,c}$ are the coefficients which differ for each output.  Then with $\bbeta_{j,l} = [\beta_{j,l,1},\dots,\beta_{j,l,C}]$ assume that
\vspace{-.2in}\beq
\bbeta_{j,l} = [\beta_{j,l,1},\dots,\beta_{j,l,C}]^T \stackrel{ind}{\sim} {\cal N}(\bzero, \bLambda_j),
\label{eq:MV_beta}
\vspace{-.1in}\eeq
i.e., multivariate normal and independent across $j$ and $l$.  

The assumption in (\ref{eq:MV_beta}) is equivalent to assuming a (separable) product correlation between the $j$-th functional components of $\eta_c$ and $\eta_{c'}$, i.e., 
\vspace{-.15in}\bdm
\Cor \left(\eta_{j,c}(\bw_j), \eta_{j,c'}(\bw'_j) \right) = \left\{
\begin{array}{ll}
\Cor \left(\eta_{j,1}(\bw_j), \eta_{j,1}(\bw'_j) \right) & \mbox{if $c = c'$} \\
\rho_{j,c,c'} \Cor \left(\eta_{j,1}(\bw_j), \eta_{j,1}(\bw'_j) \right) & \mbox{if $c \neq c'$},
\end{array}
\right.
\label{eq:MV_cor}
\vspace{-.15in}\edm
although the overall covariance structure for the entire emulator is not separable.

The multivariate discrepancy function $\bdelta$ is handled in a completely analogous manner. 
Finally, it is assumed that the multivariate error vectors are 
\vspace{-.2in}\beq
\beps_n \stackrel{iid}{\sim}, N(\bzero,\bSigma), \;\;\; n=1,\dots,N.
\vspace{-.2in}\eeq

\vspace{-.2in}
\section{Parameter Estimation}
\vspace{-.1in}
\label{sec:estimation}

As previously mentioned in Section~\ref{sec:cal_review} when the simulator is computationally demanding, it must be evaluated at some specified design points prior to the calibration procedure.  These design points $(\bx^*_1,\bt^*_1),\dots,(\bx^*_{m},\bt^*_{m})$ are generally chosen via some space filling design like Latin Hypercube Sampling (LHS) \cite{McKay79} or similar approach.  While design is also an important and active topic, it is beyond the scope of this paper to discuss it in detail.

The multivariate simulator output $\by^* = [y^*_1,\dots,y^*_C]^T$ at the $m$-th design location is
\vspace{-.1in}\beq
\by^*_m = \eta(\bx^*_m,\bt^*_m) + \bxi_m,
\label{eq:MV_sim_out}
\vspace{-.1in}\eeq
The error term $\bxi_m \stackrel{iid}{\sim} {\cal N}(\bzero,\bUpsilon)$, $m=1,\dots,M$ is included to allow for simulator ``jitter'' (e.g., solving error, or estimation error involved in the post processing of results).  \citeasnoun{Gramacy12} provide a case for the inclusion of this error, commonly called a ``nugget'', when modeling computer model output.  

Let the multivariate experimental observations $\by_n$ represented by the model in (\ref{eq:MV_data_model}) be combined with the simulator runs $\by^*_m$ represented by the model in (\ref{eq:MV_sim_out}) into one matrix of data $\bz = [\by_1,\dots,\by_N, \by^*_1,\dots,\by^*_M]$.  The goal then is to condition the unknown model parameters $\{\btheta, \delta, \eta, \bSigma, \bUpsilon\}$ on the data $\bz$.  This is accomplished with MCMC sampling.

The complete details of the MCMC algorithm, including prior specifications and full conditional distributions, etc., are provided in the Supplementary Material.  However, an overview is provided here just to illustrate the main idea.  The prior distribution for the model parameters $\btheta$ can be any distribution based on previous studies and/or expert judgment.  Section~\ref{sec:analysis} provides an example of how to choose the prior for $\btheta$.  The prior distribution of the remaining parameters amounts to specifying a prior for several covariance matrices.  If it is assumed that all of these covariance matrices are distributed as inverse-Wishart, then due to the conjugate nature of this choice, the entire MCMC procedure becomes one of conjugate Gibbs updates (with the exception of updating $\btheta$).

The MCMC routine is a typical hybrid Gibbs, Metropolis Hastings (MH) sampling scheme (see \citeasnoun{Givens05}, for example), where MH updates are performed on the elements of $\btheta$, while Gibbs updates are possible for the remaining parameters $\{\delta, \eta, \bSigma, \bUpsilon\}$.  An exception to this occurs when some of the $\theta_q$ are categorical.  In that case, the procedure can be hard to tune because the jumps between the categorical levels of $\theta_q$ cannot be made ``smaller'' to encourage higher acceptance.  However, if the entire discrepancy function $\delta$ is updated with the categorical parameters in a single MH update then better mixing can  be obtained.  The intuition being that the discrepancy function is likely to be different (i.e., multimodal) for different levels of categorical parameters.  The full conditional distribution of $\delta$ is still used in this case, but only to provide a reasonable proposal for $\delta$ conditional on the proposed values for the categorical $\theta_q$ elements.

In the calibration analysis of Section~\ref{sec:analysis}, there are also some missing data for some of the outputs.  
Fortunately, this is easily addressed within the Bayesian framework.  That is, when an element(s) of the output vector for an observation is missing, it is simply treated as another unknown parameter and sampled over in the MCMC.

As previously mentioned, a major benefit of the BSS-ANOVA approach is the scalability of the resulting MCMC algorithm.  Specifically, the algorithm is $O((J+K)(N+M))$, a result that is formalized in the Supplementary Material, immediately after the full MCMC algorithm details are provided.  Generally, the number of functional components for the emulator is $J=O((P+Q)^2)$ for a two-way interaction model or $J=O((P+Q)^2+P^2Q)$ for the limited three-way interaction model used in the results presented here.  However, as long as the dimensionality of the model input and parameter spaces are moderate relative to the number of observations and simulator runs, this approach will have a large computational advantage over the existing approaches which are $O((N+M)^3)$.

{\singlespacing
\vspace{-.2in}
\section{Calibration Analysis of the Bubbling Fluidized Bed}
\vspace{-.1in}
\label{sec:analysis}
}
In this section we provide an example calibration analysis on the bubbling fluidized bed problem which motivated the methodology described above.  First, independent prior distributions for $\btheta_q$, $q=1,\dots,6$ were obtained through a review of CFD literature on fluidized beds \cite{Li11,Chao11,Herzog12,Asegehegn11b,Hulme05,Lindborg07,Asegehegn12,Yusuf12},
\vspace{-.1in}{\small
\bdm
\theta_j \stackrel{ind}{\sim} \left\{
\begin{array}{ll}
0.1997 \: \mbox{Beta}(2.5,2.5)+0.8, & \mbox{$j=1,2$} \\
20 \: \mbox{Beta}(1.2,2.5)+25, & \mbox{$j=3,4$} \\
0.1 \: \mbox{Beta}(2.5,2.5)+0.3, & \mbox{$j=5$} \\
\mbox{Discrete Uniform$\{$'Syamlal-O'Brien', $\!$'Wen-Yu', $\!$'Gidaspow'$\!\}$}, & \mbox{$j=6$}
\end{array} \right.
\vspace{-.25in}\edm
}

Two Latin Hypercube Samples were obtained to determine where to make simulator runs.  The values of $y_1$ and $y_2$ were calculated for each simulator run at angles $\{-90.0, -67.5, -45.0, -22.5, 0.0, 22.5, 45.0, 67.5, 90.0\}^\circ$.  
Thus, the seven free ``parameters'' for which to choose values in the LHS were $(x_1, t_1, t_2, \dots, t_6)$.  However, since both $y_1$ and $y_2$ where recorded at $x_1=12.6$ cm/sec, it was decided to perform an initial LHS of 60 runs across $\btheta$ with  $x_1$ fixed at 12.6 cm/sec.  A second LHS over $\bt$ of 30 runs (10 each) at $x_1=5.5,7.0,11.0$ cm/sec, respectively, was subsequently generated since $y_1$ was recorded at these velocities as well.  In total, there are experimental observations at the four velocities (each at five distinct angles), and 90 CFD runs total covering four distinct velocities (each run provides output at 9 distinct angles for a total of $M=810$ simulator ``observations'' in the context of Section~\ref{sec:method}).  The 90 CFD runs took a total of about three weeks using Intel Xeon E5-2680 CPUs (8 cores per CFD run).

\renewcommand{\arraystretch}{.55} 
\begin{wraptable}{r}{.38\textwidth}
\vspace{-.4in}
\begin{center}
\caption{Experimental Data}
\vspace{-.1in}
\label{tab:exp_data}
\begin{tabular}{|crrcc|}
\hline
Obs & $x_1\;$ & $x_2\;$ & $y_1$ & $y_2$ \\
\hline
&&&&\\[-.12in]
1 & $\;$5.5 & -90$\;$ & 1.50 & $-$ \\
2 & $\;$5.5 & -45$\;$ & 1.70 & $-$ \\
3 & $\;$5.5 & 0$\;$ & 1.22 & $-$ \\[-.05in]
\vdots & \vdots$\;\;$  & \vdots$\;\;$ & \vdots & \vdots \\
19 & 12.6 & 45$\;$ & 0.30 & 0.07 \\
20 & 12.6 & 90$\;$ & 0.10 & 0.01 \\
\hline
\end{tabular}
\end{center}
\vspace{-.2in}
\end{wraptable}

\begin{figure}[t!]
\begin{center}
\caption{Bubble frequency from CFD simulations (lines) along with experimental data (X)}
\vspace{-.15in}
\includegraphics[width=.85\textwidth]{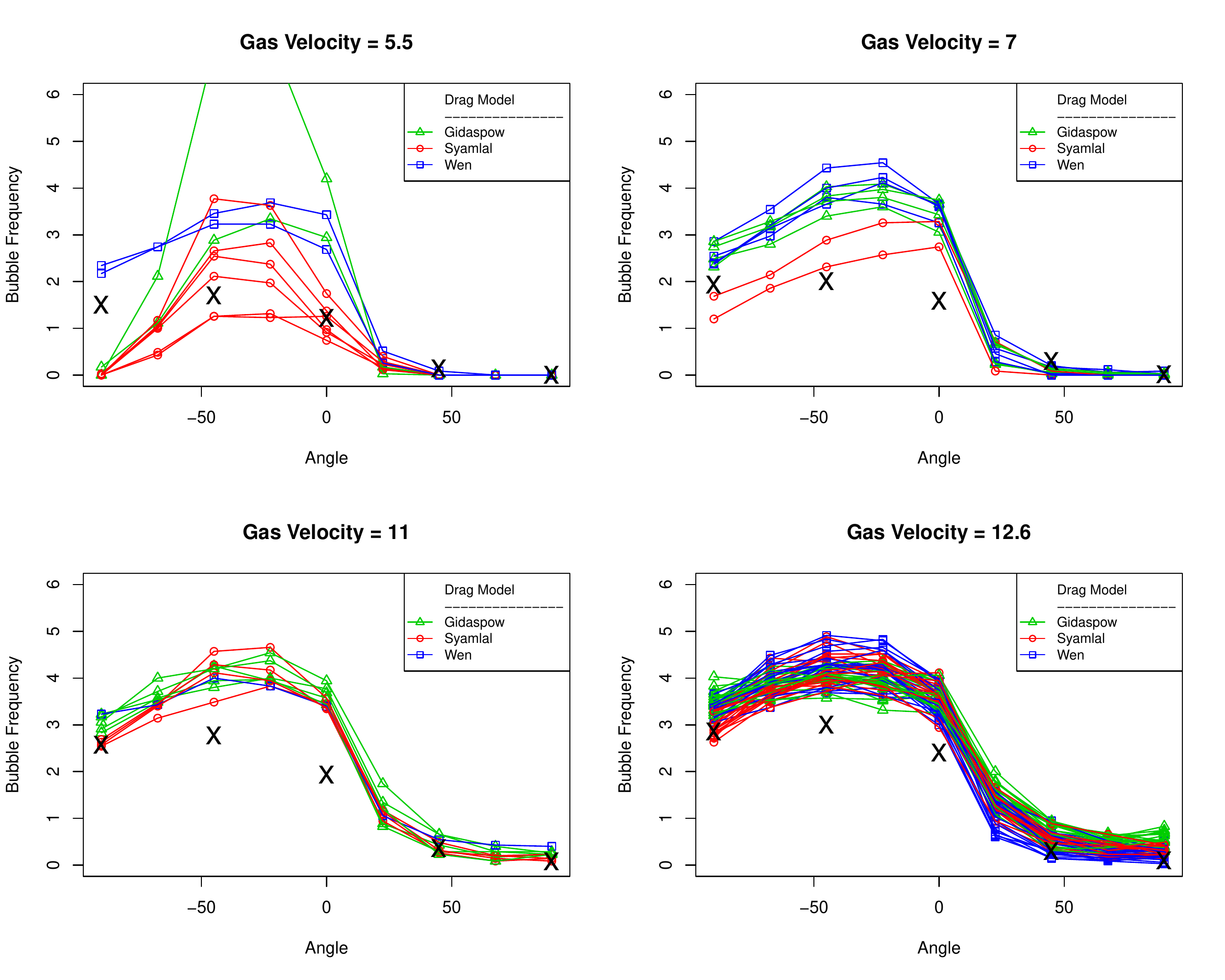}
\label{fig:bub_bed_data_1}
\end{center}
\vspace{-.2in}
\end{figure}

Table~\ref{tab:exp_data} provides a few observations of the experimental data for illustration, while Table~\ref{tab:sim_data} provides the same for a few of the simulator runs.  
There are no values for $y_2$ at velocities other than  $x_1=12.6$cm/sec.  These values were not available and were thus treated as missing data for this analysis.  
The experimental data along with the simulator runs for the output of bubble frequency ($y_1$) are displayed in Figure~\ref{fig:bub_bed_data_1}.  The full data set can be found online at the journal website.

\renewcommand{\arraystretch}{.9} 

\begin{table}[t!]
\begin{center}
\caption{Simulator Runs Data}
\vspace{-.1in}
\label{tab:sim_data}
\begin{tabular}{|ccccccccccc|}
\hline
Obs & $x_1$ & $x_2$ & $t_1$ & $t_2$ & $t_3$ & $t_4$ & $t_5$ & $t_6$ & $y^*_1$ & $y^*_2$ \\
\hline
&&&&&&&&&&\\[-.12in]
1 & 12.60 &  $\;$ 0.0 & 0.85 & 0.91 & 31.2 & 27.4 & 0.36 & Syamlal & 3.80 & 0.270 \\
2 & 12.60 & -22.5$\;$ & 0.85 & 0.91 & 31.2 & 27.4 & 0.36 & Syamlal & 4.11 & 0.344 \\[-.05in]
\vdots & \vdots & \vdots & \vdots & \vdots & \vdots & \vdots & \vdots & \vdots & \vdots & \vdots \\
808 & $\;$7.00 &  45.0 & 0.86 & 0.91 & 28.6 & 32.3 & 0.36 & Gidaspow & 0.14 & 0.002 \\
809 & $\;$7.00 &  67.5 & 0.86 & 0.91 & 28.6 & 32.3 & 0.36 & Gidaspow & 0.06 & 0.001 \\
810 & $\;$7.00 &  90.0 & 0.86 & 0.91 & 28.6 & 32.3 & 0.36 & Gidaspow & 0.03 & 0.001 \\
\hline
\end{tabular}
\vspace{-.2in}
\end{center}
\end{table}

\begin{figure}[t!]
\begin{center}
\caption{Possible variance stabilizing transformations for $y_1\!:\:$Bubble Frequency (top row) and $y_2\!:\:$Bubble Phase Fraction (bottom row), both for a gas velocity of 12.6 cm/sec.  A square root transform was used to more closely approximate the common variance assumption for $y_1$, while a logit transform was used for $y_2$.}
\label{fig:transforms}
\vspace{-.3in}
\includegraphics[width=.95\textwidth]{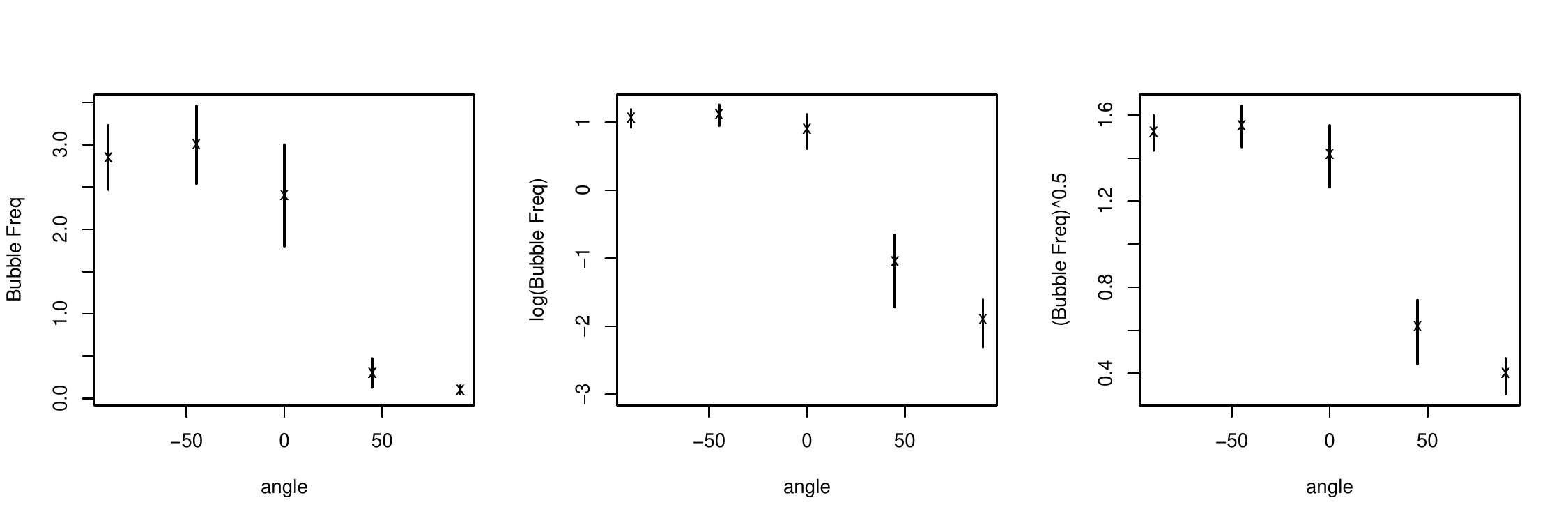}\\[-.2in]
\includegraphics[width=.95\textwidth]{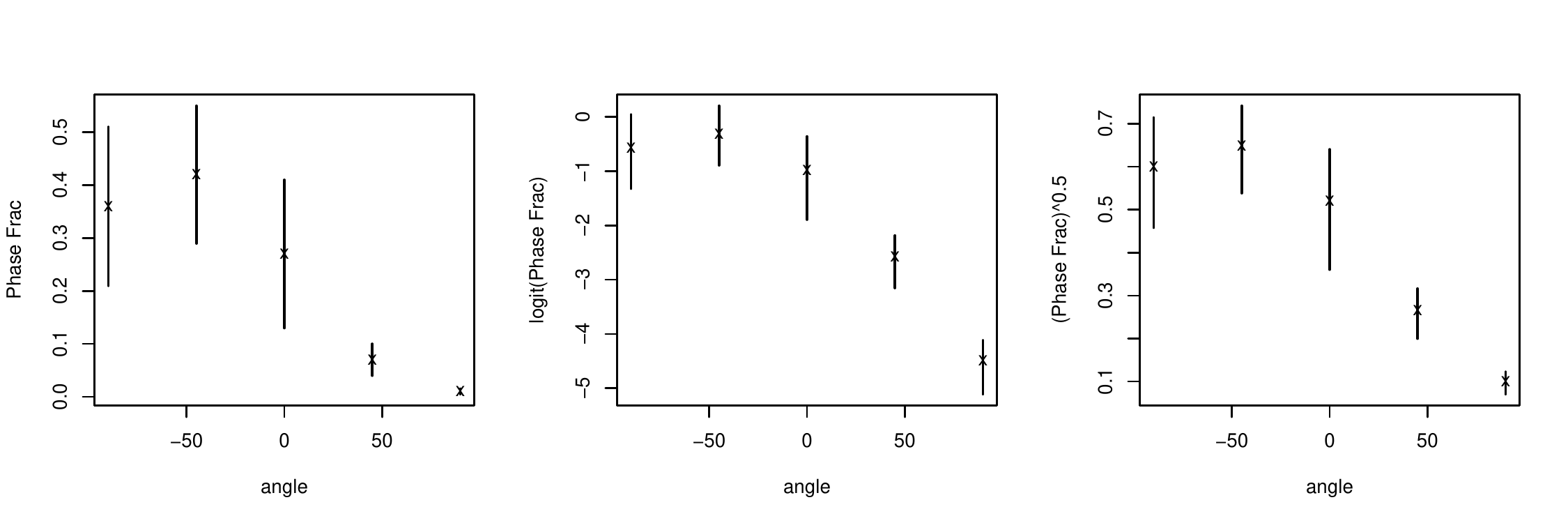}
\end{center}
\vspace{-.2in}
\end{figure}

Prior to obtaining calibration results, the commonly violated assumption in statistical models of common error variance was assessed.  As described in \citeasnoun{Lane13}, the outputs bubble frequency and bubble phase fraction are computed as time-varying averages over the course of the experiment (or simulation), once in steady state.  Standard error estimates of these averages were provided for the experimental data by \citeasnoun{Kim03}.  Figure~\ref{fig:transforms} displays $y_1$ and $y_2$ untransformed along with ($2\;\times$ standard error) bars, and the same for two possible transformations, respectively, for illustration.  The analysis presented below was conducted with the model described in Section~\ref{sec:method} applied to the transformed output vector $[\sqrt{y_1}, \;\mbox{logit}(y_2)]^T$.  These transformations were chosen via trial and error and visual inspection of plots such as those in Figure~\ref{fig:transforms}.  All of the results were transformed back to original units prior to producing any subsequent results or plots.

The standard error estimates from \citeasnoun{Kim03} were also used in the development of prior distributions for the parameters in (\ref{eq:prior_params}) of the Supplementary Material.  The use of standard errors as prior information is somewhat justified since the fluctuations in the time series of raw data are not being used as data in the analysis, just the averages. Therefore, the average of the squared standard errors (of the transformed variables) over the observations are taken to be the diagonal elements of the prior mean of both $\bSigma$ and $\bUpsilon$.  That is,
$\bSigma \sim \Wishart^{-1}(\nu_\Sigma,\bP_\Sigma)$ and $\bUpsilon \sim \Wishart^{-1}(\nu_\Upsilon,\bP_\Upsilon)$, with prior means of
\vspace{-.15in}$$
\frac{\bP_\Sigma}{\nu_\Sigma - C -1} = \frac{\bP_\Upsilon}{\nu_\Upsilon - C -1}  = \left[
\begin{array}{cc}
0.003 & 0.000 \\
0.000 & 0.009 \\
\end{array}
\right].
\vspace{-.15in}$$
The the degrees of freedom parameters were chosen to be $\nu_\Sigma = \nu_\Upsilon = 20$, and the number of output types in this problem is $C=2$.  The prior for $\bUpsilon$ was assumed to be the same as that for $\bSigma$ because the amount of time used to produce the time varying averages of the output was approximately the same for both experiment and simulation results.
The hyperpriors for $\bLambda$ and $\bOmega$ were chosen to be rather vague, specifically $(\nu_\Lambda-C-1)^{-1}\bP_\Lambda$ and $(\nu_\Omega-C-1)^{-1}\bP_\Omega$ equal to the identity matrix and $\nu_\Lambda = \nu_\Omega = 4$.

\begin{figure}[t!]
\caption{Trace plots of the model parameters from the MCMC sample.}
\vspace{-.325in}
\label{fig:trace_plots}
\begin{center}
\includegraphics[width=.85\textwidth]{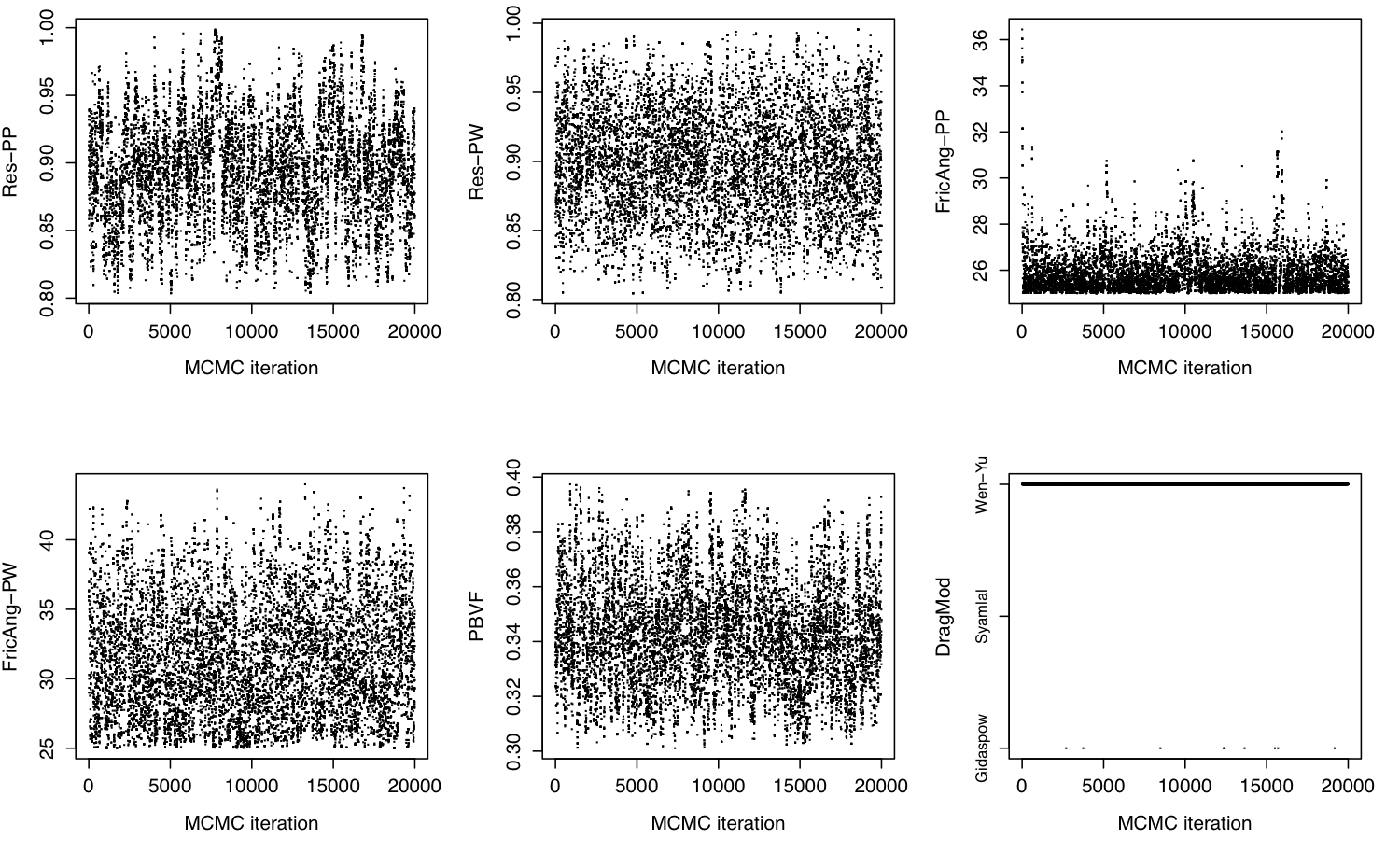}
\end{center}
\vspace{-.2in}
\end{figure}

The calibration procedure of Section~\ref{sec:method} was then applied to the data.  The MCMC routine was run for 20,000 iterations and the first 10,000 were discarded as ``burn-in''.  The 20,000 MCMC iterations with $M=810$ and $N=20$ in this case took $\sim$two hours on a MacBook Pro with a 3.06 GHz Intel Core 2 Duo.  Visual inspection of trace plots for the model parameters (Figure~\ref{fig:trace_plots}) and all variance parameters indicated that the chain appeared to be in steady state well before 10,000 iterations.  Multiple different starting points were also used to initiate 10 separate chains that all gave very similar results.  

\begin{figure}[h!]
\vspace{.1in}
\caption{Results of SA for bubble frequency across angular location for each of the four gas velocities $\{5.5, 7.0, 11.0, 12.6 \}$cm/s.}
\vspace{-.3in}
\label{fig:SA}
\begin{center}
\includegraphics[width=.85\textwidth]{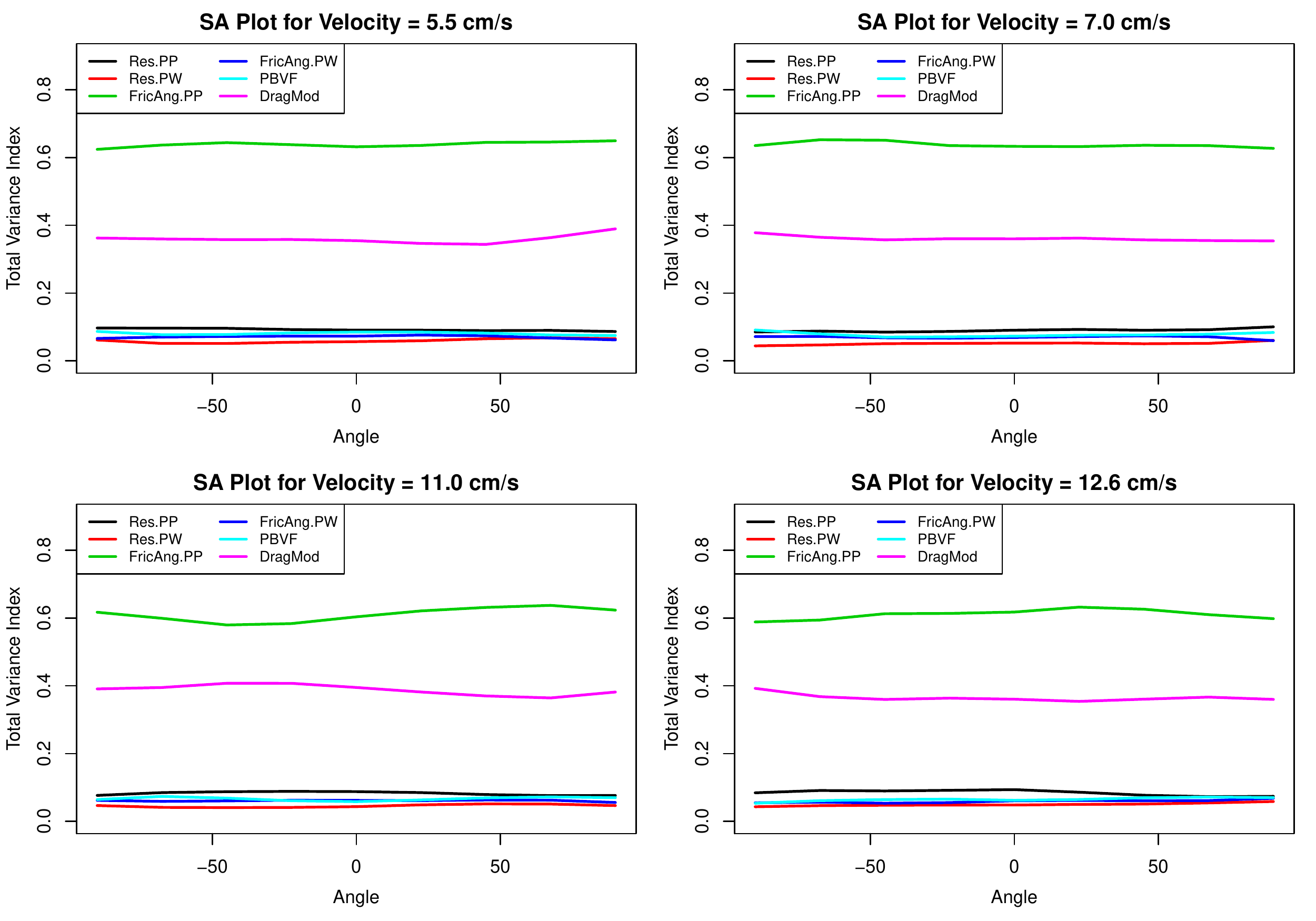}
\end{center}
\vspace{-.25in}
\end{figure}

The resulting emulator (after inverse-transforming back to the original units) from the calibration procedure was then used to perform a global sensitivity analysis (as in \citeasnoun{Storlie09b}) of the CFD model. The total variance index $T_j$ (i.e., the proportion of variance in the output that can be attributed to parameter $t_j$ and all of its interactions with other parameters; see \citeasnoun{Saltelli00}) was calculated for the effect of each of the parameters on the output Bubble Frequency across several values of the input space and is displayed in Figure~\ref{fig:SA}.  These results were obtained under the prior distribution of the model parameters (which assumed independence) as an exploratory analysis of sensitivity of the the MFIX model.  The total variance indices identify which parameters contribute significantly to the overall fluctuation of the output.  It is clear from Figure~\ref{fig:SA} that $\theta_3:$ friction angle for particle-particle interaction and $\theta_6:$ drag model are the two parameters that have the largest effect on bubble frequency.  The relative importance of the parameters is also consistent across the input space.  The same results for bubble phase fraction were obtained for velocity 12.6 cm/s as well, but not displayed. 

\begin{figure}[t!]
\begin{center}
\caption{Marginal posterior distributions for the model parameters from the MCMC sample (histograms of the y-coordinates of the points in Figure~\ref{fig:trace_plots} after 10,000 iterations) along with the prior density (blue curves).}
\vspace{-.15in}
\label{fig:posterior_theta}
\includegraphics[width=.85\textwidth]{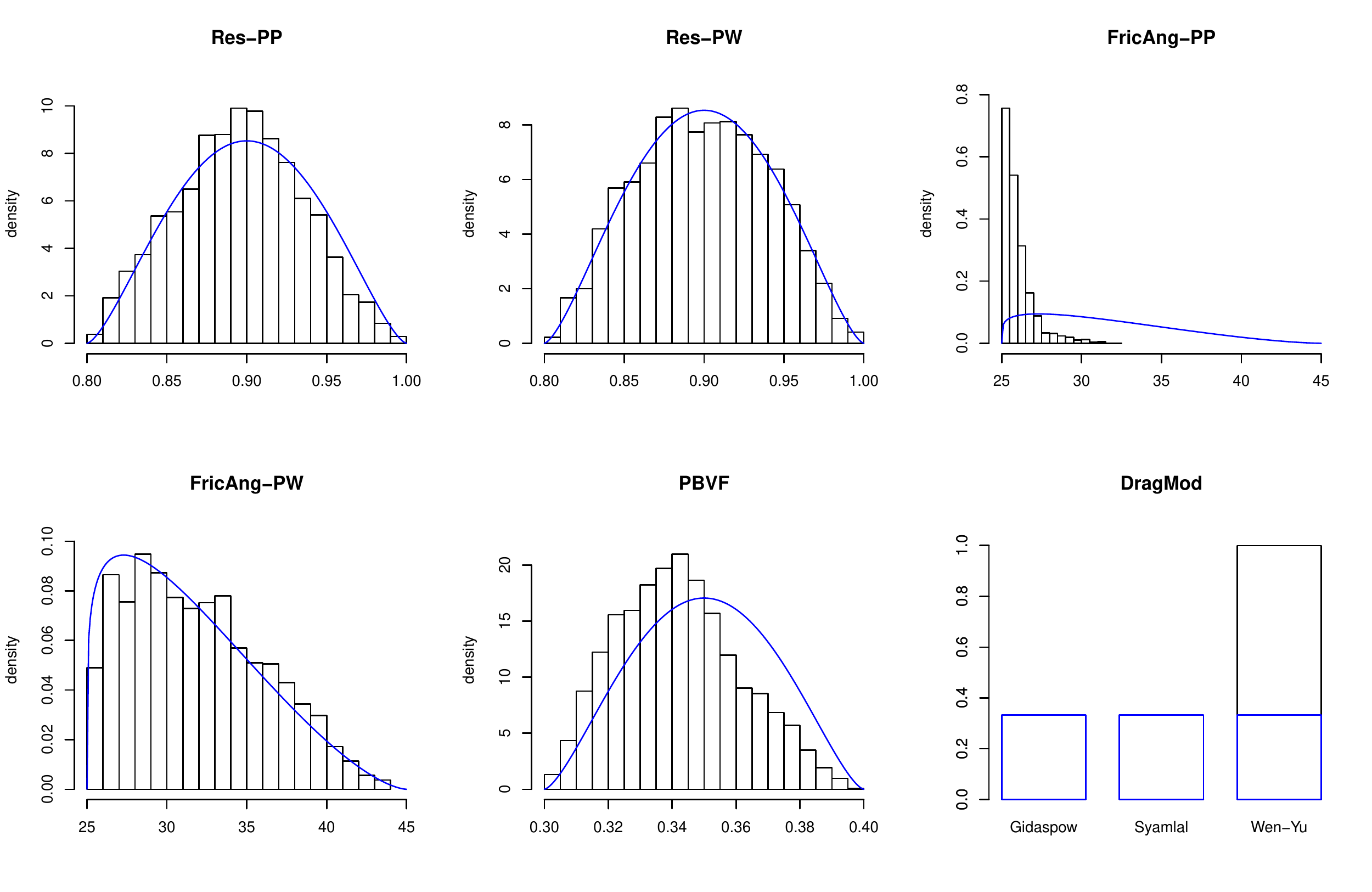}
\end{center}
\vspace{-.4in}
\end{figure}

The posterior distribution of the model parameters is displayed via marginal distributions in Figure~\ref{fig:posterior_theta}.  The marginal distributions are sufficient for illustration since there is very little correlation among the elements of $\btheta$ in the MCMC sample (the largest absolute value of posterior correlation among the $\theta_q$ was 0.07).
The Wen-Yu drag model is selected in nearly 100\% of the MCMC iterations, meaning it is the much preferred drag model for this system.  The posterior distribution of $\theta_3$, friction angle for particle-particle interaction, also changed substantially from its prior distribution.  Values of $\theta_3$ near 25$^\circ$ are much more likely given the experimental data.  The marginal posterior distributions of the other parameters are not much different from their respective priors.  This is not surprising given the lack of sensitivity of the outputs to these inputs depicted in Figure~\ref{fig:SA}.

\begin{figure}[t!]
\begin{center}
\caption{Posterior fitted plots for Bubble Frequency ($y_1$) across Angle ($x_2$) for each of the experimentally observed velocities.  The posterior mean along with 50 posterior realizations are provided for both the emulator only and the emulator plus discrepancy predictions.  The experimental data along with posterior observational error bands is also plotted.}
\vspace{-.15in}
\label{fig:fitted_y1}
\includegraphics[width=.835\textwidth]{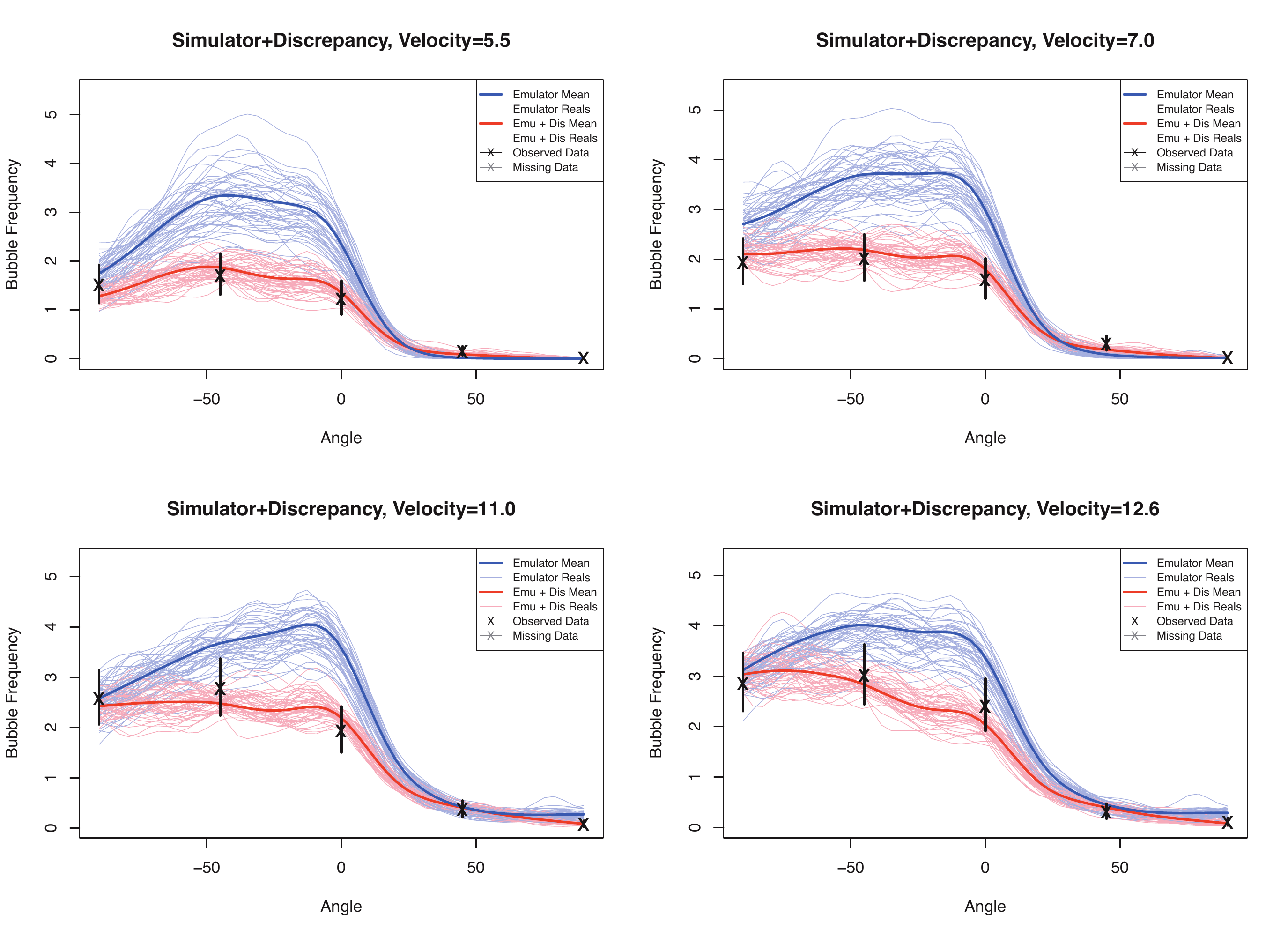}
\end{center}
\vspace{-.35in}
\end{figure}

\begin{figure}[t!]
\begin{center}
\caption{Posterior fitted plots for Phase Fraction ($y_2$) in the same format as in Figure~\ref{fig:fitted_y1}.  The uncertainty of the discrepancy is much larger for this output due to the missing observations at velocities 5.5, 7.0, and 11.0 cm/sec.}
\vspace{-.15in}
\label{fig:fitted_y2}
\includegraphics[width=.85\textwidth]{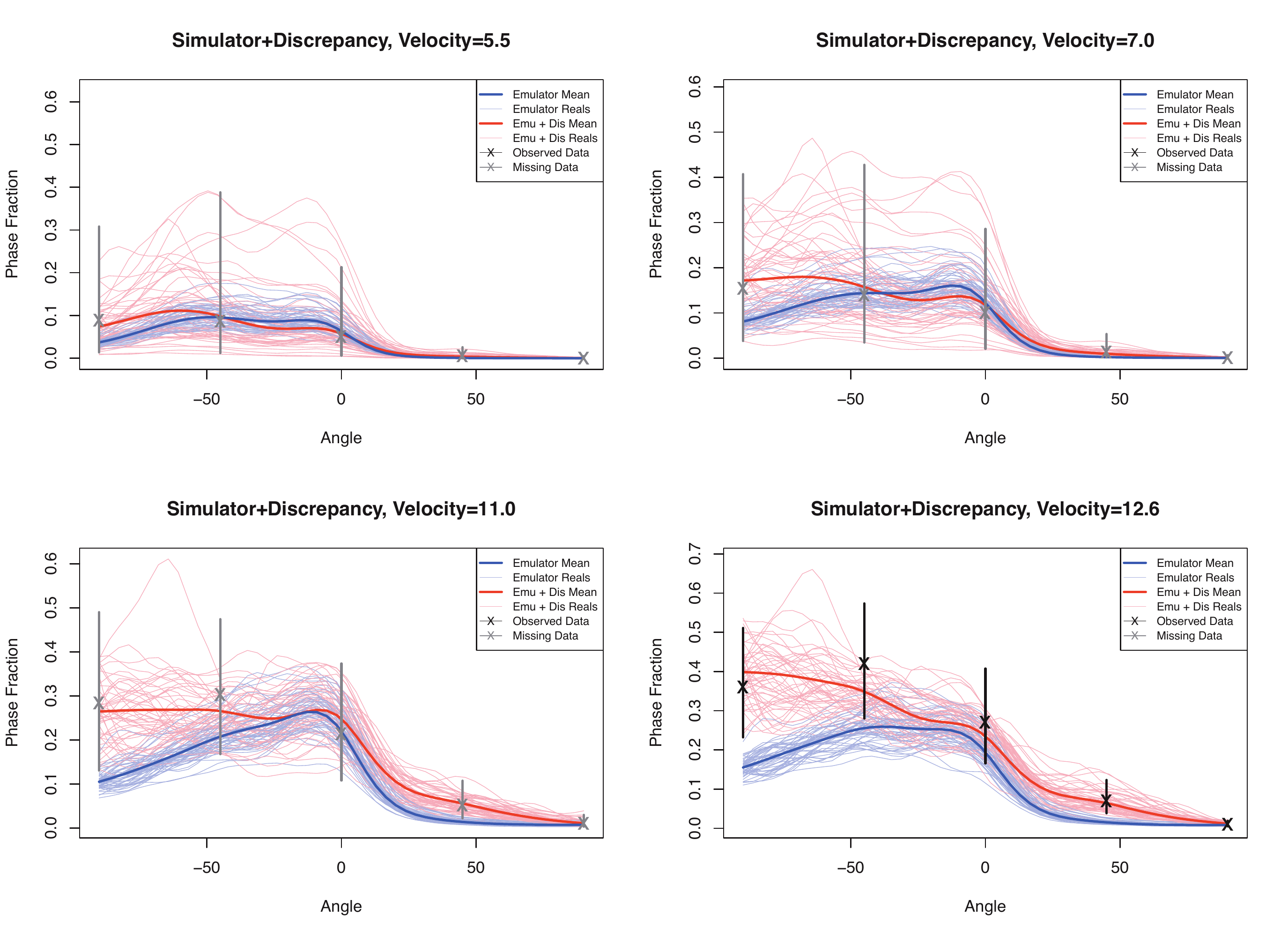}
\end{center}
\vspace{-.15in}
\end{figure}

\begin{figure}[h!]
\begin{center}
\caption{Posterior model discrepancy main effect components (posterior mean along with 50 realizations and 95\% credible bands) for $y_1:$ Bubble Frequency (row 1) and $y_2:$ Bubble Phase Fraction (row 2) across the inputs $x_1$ Velocity (column 1) and $x_2:$ Angle (column 2).}
\vspace{-.3in}
\label{fig:discrepancy}
\includegraphics[width=.85\textwidth]{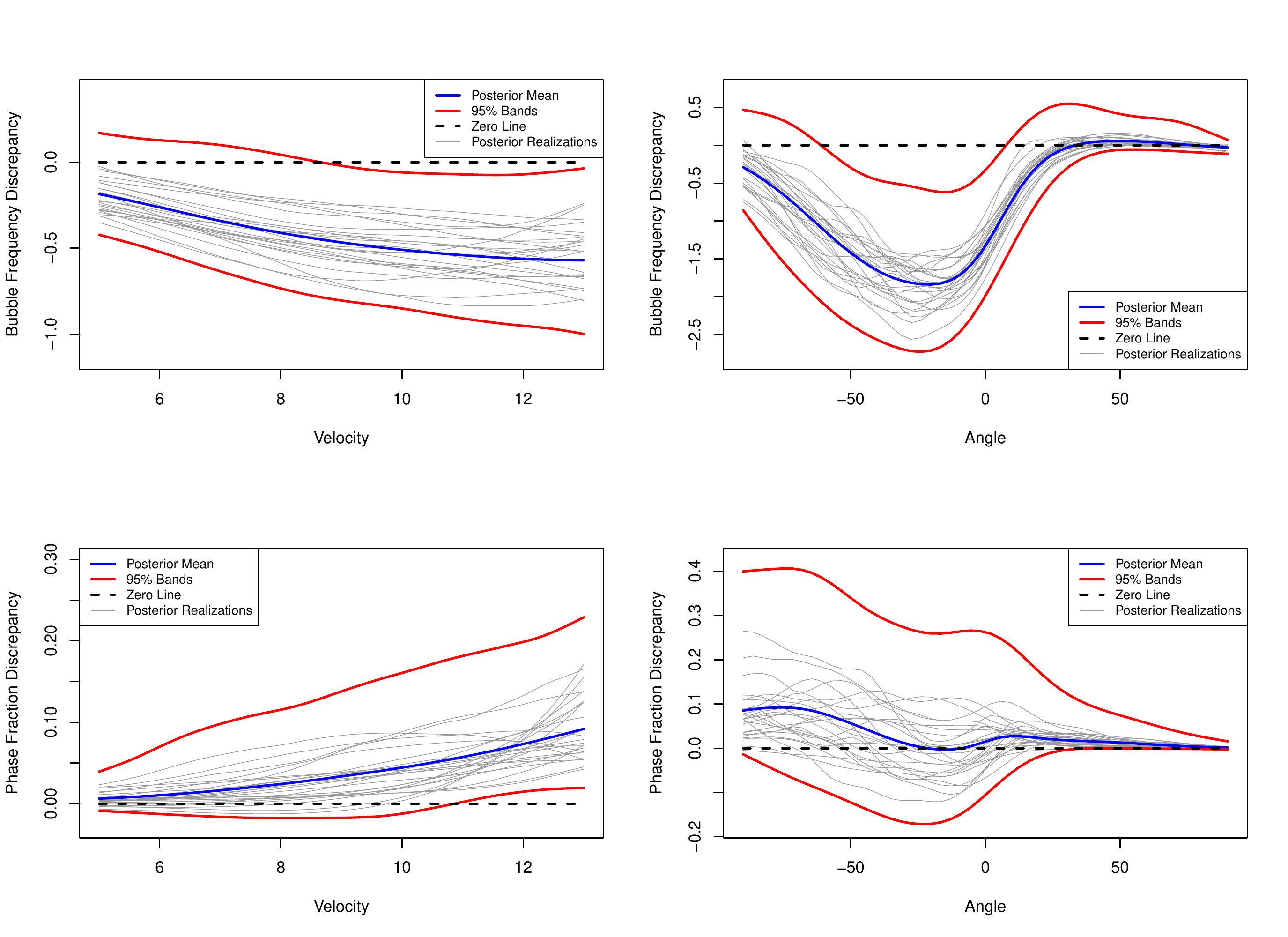}
\end{center}
\vspace{-.45in}
\end{figure}

Figures~\ref{fig:fitted_y1}~and~\ref{fig:fitted_y2} provide posterior fitted plots for the two outputs bubble frequency and bubble phase fraction, respectively.   These plots provide predictions across Angle for each observed Velocity.  Fifty prediction realizations of the simulator only (via the emulator) are provided by the blue curves.  These realizations include the uncertainty in both $\btheta$  and the emulator approximation of the simulator.  The red curves are posterior realizations of the emulator predictions plus the model discrepancy (i.e., prediction realizations of the physical system $\zeta$).  The posterior discrepancy main effects for both outputs across both of the inputs are plotted in Figure~\ref{fig:discrepancy}.  It can be seen that there is a significant non-zero discrepancy for Bubble frequency since the main effect bands across angle do not include the zero line.

\begin{figure}[t!]
\begin{center}
\caption{Out-of-sample bubble frequency predictions across angle for the four velocities (for the CDF model only and model plus discrepancy) from the calibration model (posterior mean and 95\% credible bands.  The experimental data along with posterior observational error bands is also plotted.}
\vspace{-.2in}
\label{fig:cv}
\includegraphics[width=.85\textwidth]{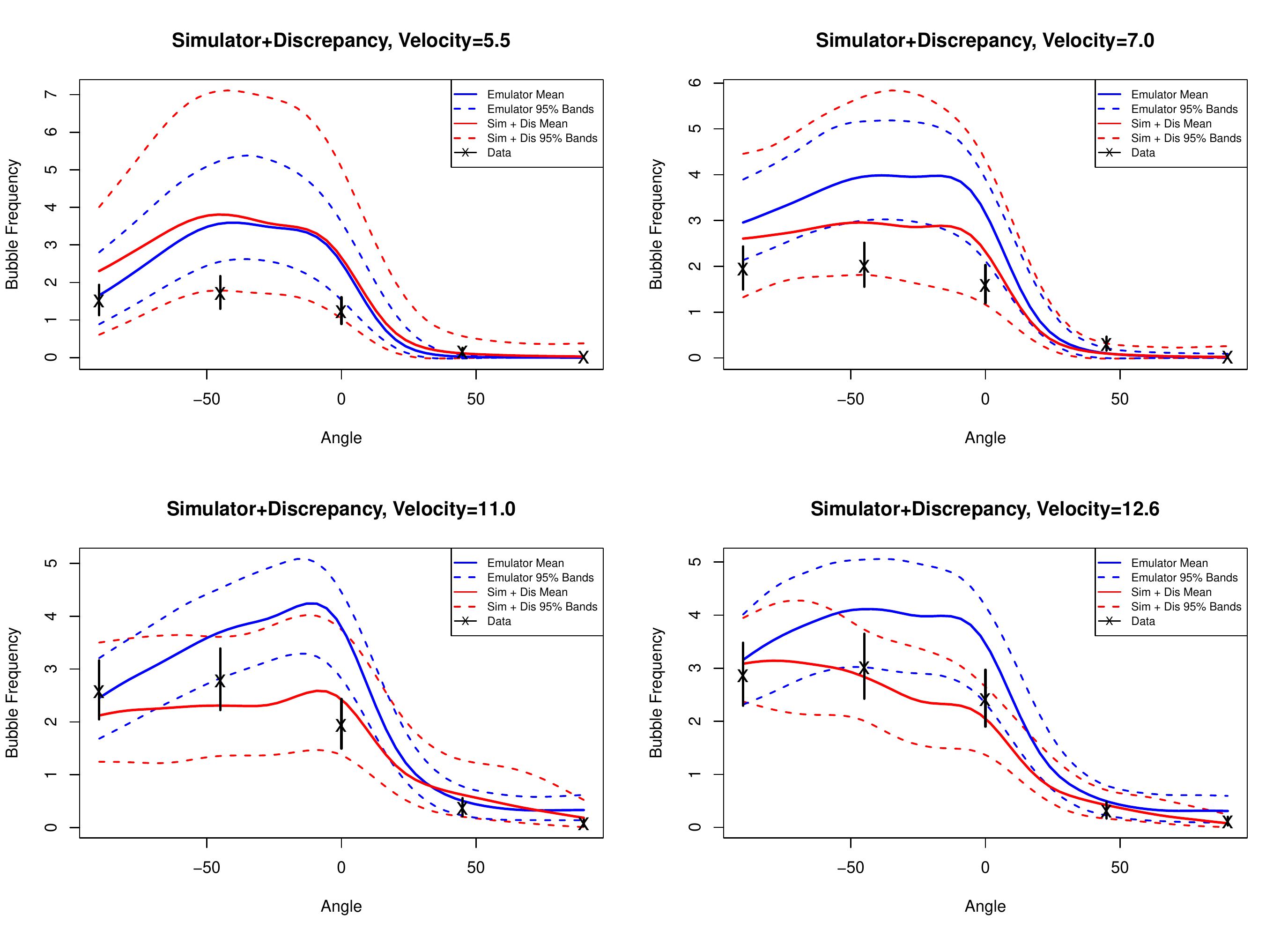}
\end{center}
\vspace{-.25in}
\end{figure}

Finally, to test the predictive accuracy of the model (with and without discrepancy), it was cross validated by removing all data from each velocity, respectively, each time estimating the parameters via MCMC, and then predicting at the removed velocity.  These out-of-sample predictions (posterior mean and 95\% credible bands) are presented in Figure~\ref{fig:cv}.  The posterior mean prediction (emulator plus discrepancy) has $R^2=0.90$ for the fitted model (in-sample), while the cross validated predictions has $R^2=0.84$ out-of-sample.  The emulator only posterior mean prediction has $R^2=0.50$ (in-sample) and cross-validated $R^2 = 0.41$.  While there is definitely some non-negligible model discrepancy in this problem, it is also clear that the CFD model captures the overall trend of the physical system across the input space reasonably well.

\vspace{-.0in}
\section{Simulation Study Results}
\label{sec:simulation}
\vspace{-.05in}

In this section, the calibration approach is used on an example that mimics the bubbling bed analysis to see exactly how well it performs on a {\em truth-known} case.  Specifically, data was generated according to the model in (\ref{eq:MV_data_model}) with exactly the same experimental input locations and LHS sample for simulator runs as that for the actual bubbling bed analysis.  The {\em synthetic} simulator $\eta$, true parameter setting $\btheta$, and discrepancy $\delta$, were taken to be the respective posterior means resulting from the bubbling bed data analysis.
Finally the error terms $\eps_i$ were generated as ${\cal N}(\bzero,\bar{\bSigma})$, where $\bar{\bSigma}$ is the posterior mean of the observational error covariance matrix resulting from the bubbling bed data analysis.  

Two simulation cases are considered. Case I is an identical setup to the problem at hand, i.e., 90 simulator runs generated via LHS in the same manner described in Section~\ref{sec:analysis} and experimental data in the same locations.  Missing values for bubble phase fraction were also assumed at velocities other than 12.6 cm/sec.  In simulation Case II, we assume that more experimental data was available (i.e., three replicates at 12 distinct velocities) and that 600 simulator runs were available.  Here we restricted the LHS sample to contain 50 runs at each of the 12 experimental velocities.

The performance of the proposed method, in terms of accuracy and computation time, is compared to that of \citeasnoun{Higdon08} which was modified to allow for categorical model parameters for this comparison.  Specifically, the isotropic correlation (a special case of \citeasnoun{Qian08}) was used in the correlation function for the categorical parameter $\theta_6$ in the emulator, and the various possible categories of $\theta_6$ were proposed in the MCMC in an independent fashion.  Because the discrepancy is not explicitly estimated in this approach (i.e., it is integrated out), this sampling strategy provided adequate mixing.  The functional principle components decomposition described in \citeasnoun{Higdon08} was applied to the output across angle ($x_2$) for this approach, which eased the computational burden.

\begin{figure}[t!]
\begin{center}
\caption{Marginal posterior distributions for the model parameters from the MCMC sample (histograms) along with the prior density (blue curves) for the first of 100 simulated data sets for simulation Case I.  True $\theta_q$ values are indicated by red tick marks.}
\vspace{-.15in}
\label{fig:theta_sim}
\includegraphics[width=.85\textwidth]{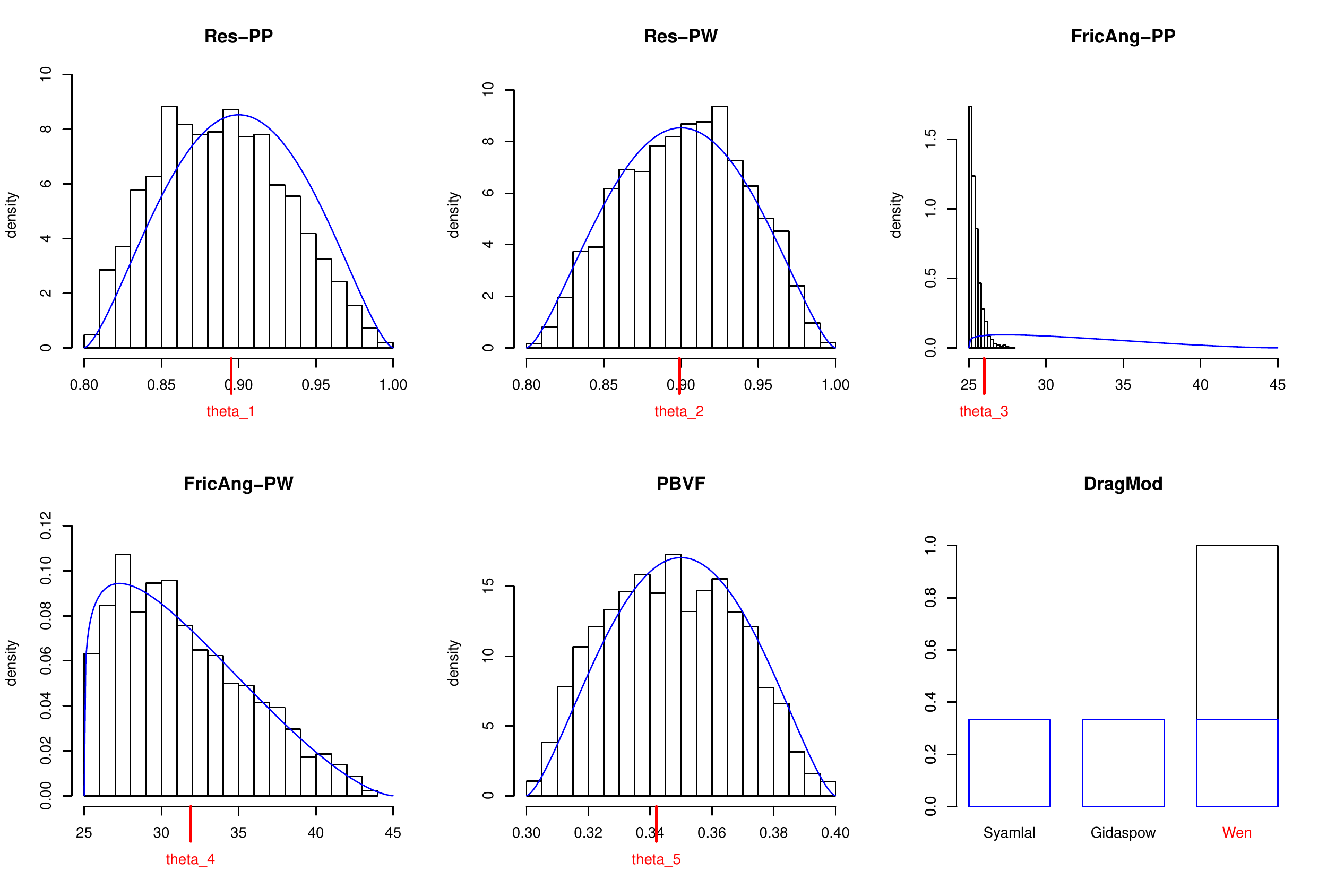}
\end{center}
\vspace{-.3in}
\end{figure}

\begin{figure}[h!]
\begin{center}
\caption{Posterior fitted plots for Bubble Frequency ($y_1$) across Angle ($x_2$) for each of the experimentally observed velocities (analogous to those in Figure~\ref{fig:fitted_y1}) for the first of 100 simulated data sets for simulation Case I.}
\vspace{-.2in}
\label{fig:fitted_sim}
\includegraphics[width=.85\textwidth]{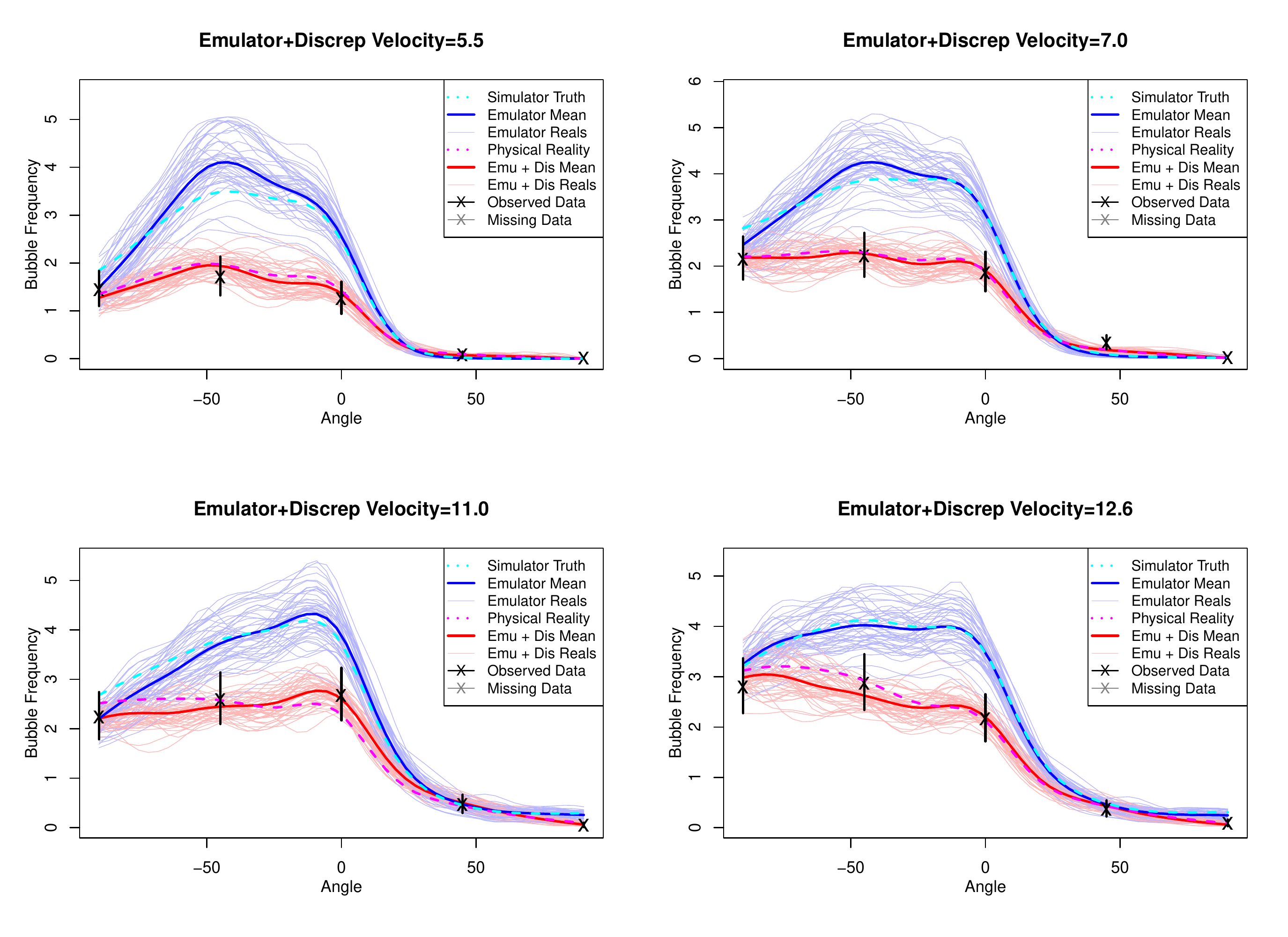}
\end{center}
\vspace{-.3in}
\end{figure}

\renewcommand{\arraystretch}{1.0} 
\begin{table}
\vspace{.1in}
\begin{center}
\caption{Simulation results for each simulation case: Average Root Posterior Mean Squared Error (ARPMSE) for continuous $\theta_q$ and the average posterior probability (APP) of the correct category for the categorical $\theta_6$.  Standard errors of theses statistics are provided in parentheses.  Computation times for 1,000 MCMC iterations are also provided.}
\label{tab:sim}
{\small
\begin{tabular}{|c||cc||cc|}
\cline{2-5}
\multicolumn{1}{c|}{} & \multicolumn{2}{c||}{\bf Case I}& \multicolumn{2}{c|}{\bf Case II} \\[.02in]
\cline{1-1} \cline{2-5}
Parameter & {\bf $\!$BSS-ANOVA} & {\bf $\;$GPMSA$\;$} & {\bf  $\!$BSS-ANOVA} & {\bf $\;$GPMSA$\;$} \\
\hline
 & \multicolumn{2}{c||}{$\;\;\;$ARPMSE} & \multicolumn{2}{c|}{$\;\;\;$ARPMSE} \\
$\theta_1$ & 0.040 (0.0004) & 0.049 (0.0007)  & 0.041 (0.0006) & 0.040 (0.0011) \\
$\theta_2$ & 0.041 (0.0005) & 0.043 (0.0004)  & 0.040 (0.0004) & 0.035 (0.0008) \\
$\theta_3$ & 0.878 (0.0256) & 2.979 (0.1666)  & 0.623 (0.0103) & 1.227 (0.0583) \\
$\theta_4$ & 4.114 (0.0146) & 4.719 (0.0741)  & 3.622 (0.0227) & 3.985 (0.0976) \\
$\theta_5$ & 0.021 (0.0002) & 0.025 (0.0004)  & 0.017 (0.0002) & 0.020 (0.0006) \\
\hline
 & \multicolumn{2}{c||}{$\;\;\;$APP}  & \multicolumn{2}{c|}{$\;\;\;$APP} \\
$\theta_6$ & 0.999 (0.0004) & 0.881 (0.0176) & 1.000 (0.0000) & 1.000 (0.0000) \\
\cline{1-1} \cline{2-5}
\cline{2-5}
\multicolumn{1}{c|}{} & \multicolumn{2}{c||}{$\;\;\;$Computation Time (min)}  & \multicolumn{2}{c|}{$\;\;\;$Computation Time (min)} \\
\multicolumn{1}{c|}{}  & 5.5 & 3.1 & 25 & 96  \\
\cline{2-5}
\end{tabular}
}
\end{center}
\vspace{-.2in}
\end{table}

Data was generated as described above to produce 100 independent data sets for each of the two simulation cases.  The estimation routine was then run on each data set and posterior results were obtained.  Figure~\ref{fig:theta_sim} provides the resulting posterior distribution of $\btheta$ from the first of the simulated data sets for case I, while Figure~\ref{fig:fitted_sim} provides a fitted plot for bubble frequency along with the ``true'' simulator and the ``true'' physical reality $\zeta$ (i.e., simulator plus discrepancy).  Table~\ref{tab:sim} summarizes the performance over all 100 data sets for each simulation case, for the proposed BSS-ANOVA approach and the modified GPMSA procedure of \citeasnoun{Higdon08}.  Table~\ref{tab:sim} reports the average root posterior mean squared error (ARPMSE) for the continuous $\theta_q$, $q=1,\dots,5$, which is the average over the results for all 100 data sets of the root posterior mean square error, $V = \{\E [ (\theta_q - \tilde{\theta}_q )^2\mid \cY ]\}^{1/2}$,
where $\tilde{\theta}_q$ is the true value of the $q$-th model parameter.  The average of the posterior probability that $\theta_6 = 3$ (i.e., the correct category) is provided to summarize estimation performance for the categorical parameter $\theta_6$.  

The two approaches provide similar estimation accuracy for both cases, with the BSS-ANOVA approach providing better estimates of the more informative parameters $\theta_3$ and $\theta_6$ (particularly in Case I).  This is to be somewhat expected since the BSS-ANOVA model form was used to generated the synthetic data.  The proposed method required slightly more time on Case I than GPMSA, but it only required 4.5$\times$ as much time for Case II than for Case I (Case II has 6.5$\times$ as large of a sample).  However, the GPMSA approach took over 30$\times$ longer on Case II than on Case I, demonstrating the better scalability of the proposed method.  The proposed approach is $4 \times$ faster than GPMSA on Case II (even with the aid of the functional principle components approach used in GPMSA).

\vspace{-.2in}
\section{Conclusions}
\vspace{-.1in}
\label{sec:conclusions}

A general approach for computer model calibration was presented for situations where there are multiple outcomes of interest, the computer model(s) are expensive to run, there are categorical parameters to calibrate, and there are missing data.  Many multiple (competing or course to fine resolution) computer model problems can also be treated with this approach.  The emulator and model discrepancy functions are represented within a Bayesian Smoothing Spline (BSS) ANOVA framework, which has linear computational complexity with the total number of data points (simulator and experimental).  

A calibration analysis of a CFD model for a bubbling fluidized bed was conducted using this approach.  It showed that low friction angles ($\sim 25^\circ$) for particle-particle interactions and the Wen-Yu drag model produce results most in line with the experimental data.  The other four model parameters investigated had very little effect on the output.  The CFD model was shown to have some discrepancy to the reality of experimental data, but also that it captured the trend of the physical reality reasonably well.

The BSS-ANOVA calibration approach compared favorably in terms of both estimation accuracy and computation time to the GPMSA approach in a simulation study that was setup to resemble the bubbling fluidized bed problem.  Still, the functional principle components approach of GPMSA is particularly appealing when the functional output is nonstationary in nature, with abrupt changes, etc.  Considering the complex nature of computer model calibration problems, it would make sense as a general rule to apply both approaches to ensure that they provide similar results.

{\singlespacing
\small
\bibliography{/Users/storlie/ncar/curt_ref.bib}
\bibliographystyle{agsm}
}

\newpage

\begin{appendix}

\setcounter{equation}{0}
\setcounter{page}{1}
\renewcommand{\theequation}{A\arabic{equation}}

\vspace{-.2in}
\begin{Large}
{\bf
Supplementary Material: ``Calibration of Computational Models with Categorical Parameters and Correlated Outputs via Bayesian Smoothing Spline ANOVA''
}
\end{Large}
\vspace{-.3in}

\section{BSS-ANOVA Basis Decomposition}

The Karhunen-Lo\'eve (KL) Theorem (\citeasnoun{Berlinet04}, pp.~65-70) guarantees that any Gaussian process $X(t)$, $t \in[a,b]$ with mean function $\mu(t)$ and continuous covariance function $K(s,t)$ can be represented as
\vspace{-.15in}
\beq
X(t) = \mu(t) + \sum_{l=1}^\infty Z_l \psi(t),
\label{eq:KL}
\vspace{-.15in}
\eeq
where (i) $Z_l \stackrel{ind}{\sim} N(0,\pi_l)$, and (ii) $\pi_l$ and $\psi_l$ are the eigenvalues and eigenfunctions of $K$, respectively, $l=1,2,\dots$.  The expression in (\ref{eq:KL}) suggests the approximation
\vspace{-.15in}
\beq
X(t) \approx \mu(t) + \sum_{l=1}^L Z_l \psi(t),
\label{eq:KL2}
\vspace{-.15in}
\eeq
for some $L$.  It is not imperative that $L$ be extremely large in practice, since it need only be large enough to allow enough high frequency eigenfunctions to represent the curve suggested by the data.


Recall, the GP model for $\eta$ in (\ref{eq:eta}), and similarly for $\delta$, is composed of many functional components, the constant $\alpha_0 \sim N(0,\varsigma_0^2)$, the main effects $\eta_r \sim GP(0,\varsigma^2_r K_1)$, and the two-way interactions $\eta_{r,r'} \sim GP(0,\varsigma^2_{r,r'} K_2)$, and similarly for higher order interactions.  The KL decomposition can conceptually be applied to each component separately.  The constant component trivially consists of one eigenfunction $\psi_1 = 1$ with eigenvalue $\varsigma_0^2$.  

The covariance function $\varsigma^2_r K_1$ for the main effect $\eta_r$ in (\ref{eq:BSS_cov}) is additive in nature, i.e., 
\beq
\varsigma^2_r K_1(s,t) = \varsigma^2_r K_{1,1}(s,t) + \varsigma^2_r K_{1,2}(s,t) + \varsigma^2_r K_{1,3}(s,t),
\label{eq:bss_cov2}
\eeq
with $K_{r,1}(s,t) = B_1(s)B_1(t)$, $K_{r,2}(s,t) = B_2(s)B_2(t)$, and $K_{r,3}(s,t) = - B_{4}(|s-t|)/4!$, so that
\vspace{-.05in}
\beq
\eta_r(t) = X_1(t) + X_2(t) + X_3(t),
\label{eq:alpha_KL}
\vspace{-.05in}
\eeq
with $X_k(t) \stackrel{ind}{\sim} GP(0,\varsigma^2_r K_{1,k})$, $k=1,2,3$.  

The first eigenvalue, eigenfunction pair for $\varsigma^2_r K_{1,1}$ are trivially $(\varsigma^2_r,B_1)$, with the rest zeros.  The same is true for $\varsigma^2_r K_{1,2}$ with $(\varsigma^2_r,B_2)$ and the rest zeros.  The number of non-degenerate eigen-pairs for $\varsigma^2_r K_{1,3}$ is not finite, and they do not have a convenient closed form, but they can be approximated with a single eigen-decomposition of the matrix $\bK_{1,3}$ resulting from evaluating $K_{1,3}$ on a dense tensor product grid (e.g., of $M=300$ equally spaced points) $\bt = [t_1,\dots,t_{M}]'$ on [0,1]).  The resulting eigen-decomposition $(\tilde{\bpi}, \tilde{\bPsi})$ of the matrix $\bK_{1,3}$, provides an approximation to the eigenvalue, eigenfunction pairs of the $K_{1,3}$ covariance function (e.g., see \citeasnoun{Ramsay05}, pp.~161, 165).  That is, the $l$-th column of $\tilde{\bPsi}$ is an approximation to the $l$-th eigenfunction $\psi_l$ of $K_{1,3}$ evaluated at each point in $\bt$, while the corresponding $l$-th element of $\tilde{\bpi}$ is an approximation to the $l$-th eigenvalue $\pi_l$ of $K_{1,3}$.  Denote these approximations as $(\tilde{\pi}_l, \tilde{\psi}_l)$, $l=1,\dots,M$.  The error of the approximation for a particular $(\pi_l, \psi_l)$ decreases as $M$ increases. 
To evaluate $\psi_l(t)$ for a point $t$ not in $\bt$, linear interpolation can be used.

Now applying the Karhunen-Lo\'eve approximation in (\ref{eq:KL}) to each $X_l$ in (\ref{eq:alpha_KL}) we obtain the result in (\ref{eq:eta_comp})
\vspace{-.05in}
\beq
\eta_r(t) \approx \sum_{l=1}^L \alpha_{r,l} \phi_l(t),
\label{eq:alpha_KL2}
\vspace{-.05in}
\eeq
with $\phi_1=B_1$, $\phi_2=B_2$, $\phi_l = \tilde{\pi}_{l-2} \psi_{l-2}$ for $l=3,\dots,L$,
and
\vspace{-.15in}
$$
\alpha_{r,l} \stackrel{iid}{\sim} N(0,\varsigma_r^2)
\vspace{-.15in}$$
Note that in the representation in (\ref{eq:alpha_KL2}), the $\phi_l$ do not depend on $r$.  That is, the basis function representation is the same for all main effects, so there is only one matrix eigen-decomposition required.  The basis function decomposition for the main effect function component for a categorical input/parameter was already explicitly constructed in the main paper in equations (\ref{eq:cat_comp}) and (\ref{eq:alpa_prior_cat}).

Two way interaction components could be obtained in the same manner (i.e., perform a matrix eigen-decomposition of a densely evaluated covariance function).  However, because of the product structure of $K_2$ in (\ref{eq:BSS_2way_cov}), the eigenfunctions will necessarily be pairwise products of the eigenfunctions from the decomposition of $K_1$ (and the same for the eigenvalues).  Therefore the two-way interactions can be written as
\vspace{-.15in}
\beq
\eta_{r,r'}(t,u) \approx \sum_{k=1}^L \sum_{l=1}^L \alpha_{r,k,l} \phi_k(t) \phi_l(u).
\label{eq:alpha_KL2-way}
\vspace{-.15in}
\eeq
Of course not all $L^2$ terms are necessary, and in practice only those terms with the top $L_{\mbox{\scriptsize 2-way}}$ eigenvalues (i.e., those with the largest values of $\int \phi_k(t)dt \int \phi_l(t)dt.$) will be kept in the actual model.  

Three-way and higher order interactions can be obtained in an entirely analogous manner.  The model used in the preceding results was a full two-way interaction model with select three-way interactions (i.e., all combinations of two input variables with a single model parameter).  For practical purposes it is important to have the matrix dimension $M \gg L$, and in the preceding analysis $M=300$, $L=25$, and $L_{\mbox{\scriptsize 2-way}}=50$ (and the number of terms used for each three-way interaction was $L_{\mbox{\scriptsize 3-way}}=100$).

Regardless of the specific functional components chosen for inclusion, the overall model for $\eta$ can now be written in general as 
\vspace{-.15in}\beq
\eta(\bw) =  \sum_{j=1}^J \sum_{l=1}^{L^\eta_j} \beta_{j,l} \varphi_{j,l}(\bw),
\label{eq:eta_sup}
\vspace{-.2in}\eeq
with $\beta_{j,l} \stackrel{ind}{\sim} {\cal N}(0,\lambda_j^2)$, and $j$ indexes over the $J$ functional components included in the model.  The $\beta_{j,l}$, $\varphi_{j,l}$, and $\lambda_j$ would correspond to a particular $\alpha_{r,l}$, $\phi_l(w_r)$, $\varsigma_r^2$, respectively, depending on the functional component represented by $j$.  The representation in (\ref{eq:eta_sup}) is for a univariate function $\eta$, but the same basis representation applies to all $C$ dimensions of a multivariate output separately.  An entirely analogous representation also holds for $\delta$.

The GP model actually used then is technically that in (\ref{eq:eta_sup}) which is simply a Bayesian linear model.  Full conditionals for all parameters have closed form conjugate distributions.  A major advantage to this computational scheme over MCMC algorithms used for GP models with traditional spatial covariance functions, is that a large matrix solve is never required for multivariate normal sampling or likelihood calculation inside of the MCMC.  The only large matrix solve required for this algorithm is that for the $M \times M$ matrix to obtain the eigen-decomposition of $K_{1,3}$, which is performed just one time prior to the MCMC iterations.

\section{MCMC Algorithm Details}
\vspace{-.05in}
\label{sec:computation}

\subsection{Linear Model Formulation of the Data Model}
Before diving into full conditionals it is prudent to first review and collate the specific details of the data model.  The  experimental observation vectors of dimension $C$ are
$$
\by_n = \eta(\bx_n, \btheta) + \delta(\bx_n) + \beps_n, \;\; n=1,\dots,N,
$$
while each of the simulator observations, also of dimension $C$ are
$$
\by^*_m = \eta(\bx^*_m, \bt^*) + \bxi_m, \;\; m=1,\dots,M.
$$
The $c$-th element of the $C \times 1$ result of $\eta(\bx, \bt)$ is
$$
\eta_c(\bx,\bt) = \beta_{0,1,c} + \sum_{j=1}^J \sum_{l=1}^{L^\eta_j} \beta_{j,l,c} \varphi_{j,l}(\bx,\bt), \;\; c=1,\dots,C,
$$
with 
\vspace{-.2in}$$
\bbeta_{j,l} = [\beta_{j,l,1},\dots,\beta_{j,l,C}]^T \stackrel{ind}{\sim} {\cal N}(\bzero, \bLambda_j), \;\; j=0,\dots,J, \; l=1,\dots,L^\eta_j.
\vspace{-.1in}$$
Similarly, the $c$-th element of the discrepancy is
$$
\delta_c(\bx) = \gamma_{0,1,c}+ \sum_{k=1}^K \sum_{l=1}^{L^\delta_k} \gamma_{k,l,c} \psi_{k,l}(\bx),\;\; c=1,\dots,C,
$$
with
\vspace{-.1in}$$
\bgamma_{k,l} = [\gamma_{k,l,1},\dots,\gamma_{k,l,C}]^T \stackrel{ind}{\sim} {\cal N}(\bzero, \bOmega_k), \;\; k=0,\dots,K, \; l=1,\dots,L^\delta_k.
\vspace{-.1in}$$

Now let $\cB_{j} = [\bbeta_{j,1}^T \mid \dots \mid \bbeta_{j,L^\eta_j}^T]^T$ be an $L^\eta_j \times C$ matrix of the regression parameters for the $j$-th functional component of the emulator, and  $\cC_{k} = [\bgamma_{k,1}^T \mid \dots \mid \bgamma_{k,L^\delta_k}^T]^T$ is an $L^\delta_k \times C$ matrix of the regression parameters for the $k$-th functional component of the discrepancy function.  Let the entire matrix of emulator coefficients be given by $\cB = [\cB_0^T \mid \cB_1^T \mid \cdots \mid \cB_J^T]^T$, and discrepancy coefficients by $\cC = [\cC_0^T \mid \cC_1^T \mid \cdots \mid \cC_K^T]^T$.

Let the design matrix for the emulator at the experimental data be given by
$$
\bPhi = [\bone_N \mid \bPhi_1 \mid \dots \mid \bPhi_J],
$$
where $\bone_N$ is a $N \times 1$ vector of ones, and $\bPhi_j$ is the $N \times L^\eta_j$ matrix
$$
\bPhi_j = \left[
\begin{array}{ccc}
\varphi_{j,1}(\bx_1,\btheta) & \cdots & \varphi_{j,L^\eta_j}(\bx_1,\btheta) \\
\vdots & \ddots & \vdots \\
\varphi_{j,1}(\bx_N,\btheta) & \cdots & \varphi_{j,L^\eta_j}(\bx_N,\btheta) \\
\end{array}
\right].
$$
Similarly define the design matrix for the discrepancy at the experimental data to be
$$
\bPsi = [\bone_N \mid \bPsi_1 \mid \dots \mid \bPsi_K],
$$
where $\bPsi_k$ is the $N \times L^\delta_k$ matrix
$$
\bPsi_k = \left[
\begin{array}{ccc}
\psi_{k,1}(\bx_1,\btheta) & \cdots & \psi_{k,L^\delta_k}(\bx_1,\btheta) \\
\vdots & \ddots & \vdots \\
\psi_{k,1}(\bx_N,\btheta) & \cdots & \psi_{k,L^\delta_k}(\bx_N,\btheta) \\
\end{array}
\right].
$$

The design matrix for the emulator at the simulator design points is
$$
\bPhi^* = [\bone_M \mid \bPhi^*_1 \mid \dots \mid \bPhi^*_J],
$$
where $\bPhi^*_j$ is the $M \times L^\eta_j$ matrix
$$
\bPhi^*_j = \left[
\begin{array}{ccc}
\varphi_{j,1}(\bx^*_1,\bt^*_1) & \cdots & \varphi_{j,L^\eta_j}(\bx^*_1,\bt^*_1) \\
\vdots & \ddots & \vdots \\
\varphi_{j,1}(\bx^*_M,\bt^*_M) & \cdots & \varphi_{j,L^\eta_j}(\bx^*_M,\bt^*_M) \\
\end{array}
\right].
$$

Now let the $N \times C$ matrix of experimental observations be
$$
\cY =  \left[\by_1^T \mid \cdots \mid \by_N^T \right]^T,
$$
and the $M \times C$ matrix of simulator observations be
$$
\cY^* = \left[{\by^*_1}^T \mid \cdots \mid {\by^*_M}^T \right]^T.
$$

Also let the $N \times C$ matrix of experimental observation errors be 
$$
\cE =  \left[\beps_1^T \mid \cdots \mid \beps_N^T \right]^T,
$$
and the $M \times C$ matrix of simulator observation errors be
$$
\cE^* =  \left[\bxi_1^T \mid \cdots \mid \bxi_M^T \right]^T.
$$

Finally, we can write the entire model for the $(N+M)\times C$ combined experimental and simulator observation matrix $\cZ$ as the multivariate linear model
\beq
\cZ = 
\left[
\begin{array}{c}
\cY \\
\cY^* \\
\end{array}
\right]
=
\left[
\begin{array}{l}
\bPhi \cB + \bPsi \cC \\
\bPhi^* \cB  \\
\end{array}
\right] 
+
\left[
\begin{array}{c}
\cE \\
\cE^* \\
\end{array}
\right].
\eeq

\subsection{Prior Specifications}

It is assumed that the experimental and simulator observation errors are, respectively,
\begin{eqnarray*}
\beps_n & \stackrel{iid}{\sim} N(\bzero, \bSigma), \;\; n=1,\dots,N \\
\bxi_m & \stackrel{iid}{\sim} N(\bzero, \bUpsilon) \;\; m=1,\dots,M.
\end{eqnarray*}
It is also assumed that the emulator and discrepancy coefficients are, respectively,
\begin{eqnarray*}
\vec{\cB}_j & \stackrel{ind}{\sim} & N(\bzero, \bLambda_j \otimes \bI_{L^\eta_j}), \;\; j=0,\dots,J \\
\vec{\cC}_k & \stackrel{ind}{\sim} & N(\bzero, \bOmega_k \otimes \bI_{L^\delta_k}), \;\; k=0,\dots,K,
\end{eqnarray*}
where $\otimes$ is the Kronecker product, $\bI_L$ is the $L \times L$ identity matrix, and $\vec{\bA}$ is the vectorization of a matrix $\bA$.  That is, for a $L \times C$ matrix $\bA$, $\vec{\bA}$ is the $LC \times 1$ column vector obtained by stacking the columns of the matrix $\bA$ on top of one another.

Thus, the model specified for the function $\eta$ requires only a prior specification for $\bLambda_j$, $j=1,\dots,J$.  Similarly, the model for $\delta$ requires only a prior specification for $\bOmega_k$, $k=1,\dots,K$.  Therefore the full list of parameters needing a prior specification is 
\vspace{-.1in}\beq
\{\btheta, \bLambda_1, \dots, \bLambda_J, \bOmega_1, \dots, \bOmega_K, \bSigma, \bUpsilon\}.
\label{eq:unknown_params}  
\vspace{-.1in}\eeq
Below we describe the structure of the prior specification for each of the elements of the list in (\ref{eq:unknown_params}), each of which is assumed independent of the other elements.

As mentioned in the main paper, the prior distribution for the model parameters $\btheta$ is allowed to be any distribution based on previous studies and/or expert judgment.  Section~\ref{sec:analysis} provided an example of how to choose the prior for $\btheta$.  The remaining parameters in (\ref{eq:unknown_params}) are covariance matrices.  If it is assumed that all of these matrices are distributed as inverse-Wishart, then due to the conjugate nature of this choice, the entire MCMC procedure (with the exception of updating $\btheta$) becomes Gibbs sampling.  The Wishart is sufficiently flexible to incorporate prior belief in most cases.  There may be exceptions to this and in such cases another distribution can be used, however the MCMC routine will not be as clean (i.e., require more tuning and care to ensure good mixing, etc.).

Thus, for the remainder of the discussion, it is assumed that
\vspace{-.2in}\begin{eqnarray}
\bLambda_j^{-1} & \stackrel{ind}{\sim} & \Wishart(\bP_\Lambda, \nu_\Lambda), \;\; j=0,\dots,J\\[-.075in]
\label{eq:Lambda_prior}
\bOmega_k^{-1} & \stackrel{ind}{\sim} & \Wishart(\bP_\Omega, \nu_\Omega), \;\; k=0,\dots,K\\[-.075in]
\label{eq:Omega_prior}
\bSigma^{-1} & {\sim} & \Wishart(\bP_\Sigma, \nu_\Sigma), \\[-.075in]
\label{eq:Sigma_prior}
\bUpsilon^{-1} & {\sim} & \Wishart(\bP_\Upsilon, \nu_\Upsilon).\\[-.55in] \nonumber
\label{eq:Upsilon_prior}
\end{eqnarray}
The Wishart$(\bP,\nu)$  distribution is a multivariate generalization of the Chi-squared distribution, with the mean equal to $\nu \bP$ (a $C \times C$ matrix in this case), and $\nu$ (a scalar) is the degrees of freedom parameter.
Therefore to complete the prior specification, a distribution must be provided for $\btheta$, and values must be provided for 
\vspace{-.2in}\beq
\left\{\bP_\Lambda, \nu_\Lambda, \bP_\Omega, \nu_\Omega, \bP_\Sigma, \nu_\Sigma, \bP_\Upsilon, \nu_\Upsilon \right\}.
\label{eq:prior_params}
\vspace{-.2in}\eeq
Section~\ref{sec:analysis} discussed the selection of these values in the context of the bubbling bed problem.

\subsection{MCMC Algorithm and Full Conditional Distributions}
\label{sec:full_cond}

The entire collection of parameters to be sampled in the MCMC is
\vspace{-.2in}\beq
\Theta = \left\{\btheta, \{\cB_{j}\}_{j=0}^J, \{\cC_{k}\}_{k=0}^K, \{\bLambda_j\}_{j=0}^J, \{\bOmega_k\}_{k=0}^K \bUpsilon, \bSigma, \{\by_n^{\miss}\}_{n=1}^N \right\},
\label{eq:param_set}
\vspace{-.2in}
\eeq
where $\by_n^{\miss}$ is a vector containing any missing elements from the $n$-th experimental observation vector $\by_n$.  The MCMC algorithm proceeds with Gibbs updates for each of the elements of $\Theta$ with the exception of $\btheta$, the elements of which $\theta_q$, $q=1,\dots,Q$ are updated via a Metropolis Hastings (MH) step.  Full conditional distributions with which to perform the Gibbs updates are provided below for all of the parameter groups listed in (\ref{eq:param_set}) except for $\btheta$, in which case the specifics of the MH step is described instead.

\noindent
{$\underline{\bLambda_j \mid \mbox{rest}}$}

\noindent
Conditional on all other parameters and the data (i.e., rest), $\bLambda_j$ only depends on $\cB_j$. This reduces to a multivariate normal with mean known (to be zero) and covariance matrix unknown with an inverse-Wishart prior distribution.  It is a well known result that the inverse-Wishart is conjugate for the covariance matrix in this case and therefore
$$
\bLambda_j \mid \mbox{rest} \sim \Wishart^{-1}(\bP^*, \nu^*),
$$ where $\bP^* = \cB_j^T \cB_j + \bP_\Lambda$ and $\nu^* = L^\eta_j + \nu_\Lambda$, for $j=0,\dots,J$.\\[-.1in]

\noindent
{$\underline{\bOmega_k \mid \mbox{rest}}$}

\noindent
In an analogous manner $\bOmega_k$ conditional on all other parameters and the data, only depends on $\cC_j$, and due to conjugacy of the Wishart
$$\bOmega_k \mid \mbox{rest} \sim \Wishart^{-1}(\bP^*, \nu^*),$$
 where $\bP^* = \cC_j^T \cC_j + \bP_\Omega$ and $\nu^* = L^\delta_k + \nu_\Omega$, for $k=0,\dots,K$.\\[-.1in]

\noindent
{$\underline{\bSigma \mid \mbox{rest}}$}

\noindent
In a similar manner again, $\bSigma$ conditional all other parameters and the data reduces to a case where the experimental data residuals are known, and they are multivariate normal with mean known (to be zero) and covariance ($\bSigma$) unknown but with an inverse-Wishart prior.  Thus
$$
\bSigma_k \mid \mbox{rest} \sim \Wishart^{-1}(\bP^*, \nu^*),
$$ where $\bP^* = \cE^T \cE + \bP_\Sigma$ and $\nu^* = N + \nu_\Sigma$.\\[-.1in]

\noindent
{$\underline{\bUpsilon \mid \mbox{rest}}$}

\noindent
By the same logic as that for $\bSigma \mid \mbox{rest}$, now with the simulator residuals,
$$
\bUpsilon_k \mid \mbox{rest} \sim \Wishart^{-1}(\bP^*, \nu^*),$$ 
where $\bP^* = {\cE^*}^T \cE^* + \bP_\Upsilon$ and $\nu^* = M + \nu_\Upsilon$.\\[-.1in]

\noindent
{$\underline{\cC_k \mid \mbox{rest}}$}

\noindent
Define the remainder of the observation matrix $\cY$ after subtracting off all terms in the model but the error and the $k$-th functional component of the discrepancy as
\beq
\cY_{(-k)} = \cY - \bPhi \cB - \bPsi_{(-k)} \cC_{(-k)} = \bPsi_k \cC_k + \cE.
\label{eq:mat_C_update}
\eeq
The matrix relation in (\ref{eq:mat_C_update}) can be written in vectorized form as 
$$
\vec{\cY}_{(-k)} = \left( \bI_C \otimes \bPsi_k \right) \vec{\cC}_k + \vec{\cE}.
$$
Therefore $\cC_k$ conditional on all other parameters and the data reduces to a linear model with covariance of the errors known and normal prior on the coefficients $\vec{\cC}_k$.  The normal distribution is well known to be conjugate in this setting \cite{Gelman03} and thus
\bdm
\vec{\cC}_k \mid \mbox{rest} \sim N(\bm^*, \bV^*)
\edm 
where
\begin{eqnarray*}
\bV^* & = & \left[ \left(\bI_C \otimes \bPsi_k \right)^T \left(\bSigma \otimes \bI_N \right)^{-1} \left(\bI_C \otimes \bPsi_k \right) + \left(\bOmega_k \otimes \bI_{L^\eta_j} \right)^{-1} \right]^{-1} \\
&=& \left[ \bSigma^{-1} \otimes  \bPsi_k^T \bPsi_k + \bOmega_k^{-1} \otimes \bI_{L^\eta_j} \right]^{-1}
\end{eqnarray*}
and
\begin{eqnarray*}
\bm^* & = & \bV^* \left[ \left( \bI_C \otimes \bPsi_k \right)^T \left(\bSigma \otimes \bI_N \right)^{-1} \vec{\cY}_{(-k)} \right] \\
&=& \bV^* \left(\bSigma^{-1} \otimes  \bPsi_k^T \right) \vec{\cY}_{(-k)} 
\end{eqnarray*}

\noindent
{$\underline{\cB_j \mid \mbox{rest}}$}

\noindent
The setup here is similar to that for $\cC_k \mid \mbox{rest}$ only that the remainder of $\cY^*$ in addition to that of $\cY$ needs to be considered because the $\cB_k$ are also dependent on the simulator runs.  Thus, define the remainder
\beq
\cZ_{(-j)} = \left(
\begin{array}{c}
\cY_{(-j)} \\
\cY^{\;*}_{(-j)} \\
\end{array}
\right)
=
\left(
\begin{array}{l}
\cY - \bPhi_{(-j)} \cB_{(-j)} - \bPsi \cC \\
\cY^* - \bPhi_{(-j)}^* \cB_{(-j)} 
\end{array}
\right)
= 
\left(
\begin{array}{c}
\bPhi_j \cB_j  \\
\bPhi^*_j \cB_j
\end{array}
\right)+
\left( 
\begin{array}{c}
{\cE} \\
{\cE^*}
\end{array}
\right).
\label{eq:mat_B_update}
\eeq
The matrix relation in (\ref{eq:mat_B_update}) can be written in vectorized form as 
$$
\vec{\cZ}_{(-j)} = \left( 
\begin{array}{c}
\left( \bI_C \otimes \bPhi_j \right) \vec{\cB}_j \\
\left( \bI_C \otimes \bPhi^*_j \right) \vec{\cB}_j + \vec{\cE^*}
\end{array}
\right) + 
\left( 
\begin{array}{c}
\vec{\cE} \\
\vec{\cE^*}
\end{array}
\right).
$$
Therefore $\cB_k$ conditional on all other parameters and the data reduces to a linear model with covariance of the errors known and normal prior on the coefficients $\vec{\cB}_k$.  Again, the normal distribution is conjugate for $\vec{\cB}_k$ so 
\beq
\vec{\cB}_k \mid \mbox{rest} \sim N(\bm^*, \bV^*)
\label{eq:beta_update}
\eeq
where
\begin{eqnarray*}
\bV^* & = & \left[ 
\left(\bI_C \otimes \bPhi_j^T \mid \bI_C \otimes {\bPhi^*_j}^T \right)
\left( 
\begin{array}{cc}
\bSigma^{-1} \otimes \bI_N & \bzero \\
\bzero & \bUpsilon^{-1} \otimes \bI_M
\end{array}
\right)
\left(
\begin{array}{c}
\bI_C \otimes \bPhi_j \\
\bI_C \otimes {\bPhi^*_j}
\end{array}
\right)
+
\bLambda_j^{-1} \otimes \bI_{L^\eta_j}
 \right]^{-1} \\
&=& \left[ 
\bSigma^{-1} \otimes \left( \bPhi_j^T\bPhi_j \right) + \bUpsilon^{-1} \otimes \left(  {\bPhi_j^*}^T\bPhi_j^* \right) +  \bLambda_j^{-1} \otimes \bI_{L^\eta_j}
\right]^{-1}
\end{eqnarray*}
and
\begin{eqnarray*}
\bm^* & = & \bV^* \left[
\left(\bI_C \otimes \bPhi_j^T \mid \bI_C \otimes {\bPhi^*_j}^T \right)
\left( 
\begin{array}{cc}
\bSigma^{-1} \otimes \bI_N & \bzero \\
\bzero & \bUpsilon^{-1} \otimes \bI_M
\end{array}
\right)
\left(
\begin{array}{c}
\vec{\cY}_{(-j)} \\
\vec{\cY}^{\;*}_{(-j)} 
\end{array}
\right) 
\right] \\
&=& \bV^* \left[  
\left( \bSigma^{-1} \otimes \bPhi_j^T \right) \vec{\cY}_{(-j)} + \left( \bUpsilon^{-1} \otimes {\bPhi_j^*}^T \right) \vec{\cY}^{\;*}_{(-j)}
\right]
\end{eqnarray*}

\noindent
{$\underline{\by_n^{\miss} \mid \mbox{rest}}$}

\noindent
Conditional on all of the parameters, $\by_n$ and $\beps_n$ contain equivalent information.  Let $\beps_n^{\miss}$ contain the elements of $\beps_n$ corresponding the the missing elements of $\by$ and let $\beps_n^{\obs}$ contain the elements of $\beps_n$ corresponding the the observed elements of $\by$.  It was assumed that $\beps \sim N(\bzero, \bSigma)$, so equivalently after reordering the rows of $\beps$ so that all missing values are together, we can write
$$
\left(
\begin{array}{c}
\beps_n^{\miss}\\
\beps_n^{\obs}
\end{array}
\right)
\sim
N\left( \bzero, \left(
\begin{array}{cc}
\bSigma_{11} & \bSigma_{12}\\
\bSigma_{12}^T & \bSigma_{22}
\end{array}
\right)
\right).
$$
Combining this and the fact that $\beps_n$ are independent across $n$, $\by_n^{\miss} \mid \mbox{rest}$ reduces to 
$$
\beps_n^{\miss} \mid \beps_n^{\obs} \sim N(\bSigma_{12} \bSigma_{22}^{-1}\;,\; 
\bSigma_{11} - \bSigma_{12}\bSigma_{22}^{-1}\bSigma_{12}^T).
$$

\noindent
{\underline{MH update for $\theta_q$}}

\noindent
The full conditional distribution of $\theta_q \mid \mbox{rest}$ does not have a convenient form with which to perform Gibbs updates.  However, the MH ratio has a very simple form which is easy to compute.  Once again, conditional on all parameters the experimental data $\cY$ is equivalent to $\cE$, and similarly $\cY^*$ is equivalent to $\cE^*$.  The entire data likelihood is then
\beq
\cL(\cY, \cY^* ; \Theta) = \cL_1(\cY^* ; \Theta) \cL_2(\cY^* ; \Theta) = 
\left[\prod_{n=1}^N \cN \left(\beps_n; \bzero, \bSigma \right)\right] 
\left[\prod_{m=1}^M \cN \left(\bxi_m; \bzero, \bUpsilon \right)\right] 
\label{eq:theta_like}
\eeq
where $\cN \left(\;\cdot\;; \bzero, \bSigma \right)$ is the multivariate normal density with mean $\bzero$ and covariance matrix $\bSigma$.  Suppose the proposed move from $\theta_q$ to $\theta_q^*$ is governed by the density $d(\theta_q^* \mid \theta_q)$.  Also let $\Theta^*$ denote the set of parameters $\Theta$ that has $\theta_q$ replaced by $\theta_q^*$.  The MH ratio is for a proposal $\theta_q^*$ is then
\beq
MH = \frac{\cL(\cY, \cY^* ; \Theta^*) \pi(\Theta^*) d(\theta_q \mid \theta_q^*)}
{\cL(\cY, \cY^* ; \Theta) \pi(\Theta) d(\theta_q^* \mid \theta_q)}.
\label{eq:MH_1}
\eeq
However, the $\cL_2$ portion of the likelihood in (\ref{eq:theta_like}) remains unchanged regardless of the value of $\theta_q$.  Also, the prior for $\btheta$ is assumed independent of the other parameters resulting in $\pi(\Theta) = \pi(\btheta)\pi(\Theta \setminus \btheta)$, where $\pi(\Theta \setminus \btheta)$ remains unchanged regardless of the value of $\theta_q$.  Therefore (\ref{eq:MH_1}) can be reduced to
$$
MH = \frac{\cL_1(\cY ; \Theta^*) \pi(\btheta^*) d(\theta_q \mid \theta_q^*)}
{\cL_1(\cY ; \Theta) \pi(\btheta) d(\theta_q^* \mid \theta_q)},
$$
which only requires $2N$ evaluations of a multivariate normal density of dimension $C$, two evaluations of the proposal density for $\theta_q$, and two evaluations of the prior density for $\btheta$.

The proposal used to produce the results presented in the paper was a logit normal distribution with parameters $\mu^* = \logit(\theta_q)$ and $\sigma^*$ a tuning parameter set to encourage $\sim\!30$\% acceptance.  An exception occurs when some of the $\theta_q$ are categorical.  In this case a single MH update is proposed for all of the categorical elements of $\btheta$ and for the coefficients of the discrepancy $\cC$.  The categorical $\theta_q$ are proposed equally likely for each category and $\cC^*$ are proposed using the full conditionals of $\cC_k \mid$ rest for $k=1,\dots,K$.  That is, a proposal is constructed for each of the categorical $\theta_q$, then at each $k$ in the proposal formulation of $\cC_k$, the current $\cC_k$ value is replaced with $\cC_k^*$ for the proposal of the further $\cC_{k'}$ for $k'>k$ in the conditional rest (i.e., all other parameters) set.  The proposal density is then evaluated accordingly.

\subsection{Computational Complexity of the MCMC Algorithm}

Unlike other approaches to computer model calibration in the literature which are based on the squared exponential GP emulation and are generally $O\{(N+M)^3\}$ for each MCMC iteration, the proposed BSS-ANOVA approach is linear in $N+M$.  There is no free lunch, however, so to be fair there is additional complexity in the proposed approach that is not part of the traditional GP approach.  Specifically, the number of functional components for emulator and discrepancy ($J$ and $K$, respectively) now factor into the computational complexity.  

From the results of Section~\ref{sec:full_cond}, the update of $\bLambda_j$, requires the construction of $\bP^* = \cB_j^T \cB_j + \bP_\Lambda$ which is $O(L^\eta_j C^2)$
and the random generation of a $C \times C$ inverse-Wishart which is $O(C^3)$.   Recall that $C$ is the number of dimensions of the output vector $\by$.  The entire update is then $O(L^\eta_j C^2 + C^3)$, and there are a total of $J$ such updates.  Similarly each of the $K$ updates of $\bOmega_k$ is $O(L^\delta_k C^2 + C^3)$.  The single updates for $\bUpsilon$, and $\bSigma$ are $O(NC^2+C^3)$ and $O(MC^2+C^3)$, repsectively.  Let $L=\max\{L^\eta_j,L^\delta_k\}$.  The updating of this entire block of parameters is then 
\beq
O\{(L(J+K)+N+M)C^2 + (J+K)C^3\}.
\label{eq:order_var}
\eeq

Updates of $\by_n^{\miss}$ are less than $O(C^3)$ and there are at most $N$ total updates, resulting in a complexity less than $O(NC^3)$.  The complexity of each $\theta_q$ update is essentially the evaluation of the experimental data likelihood (evaluation of prior density and proposal density are $O(1)$).  The likelihood evaluation is also $O(NC^3)$ and thus the updating of $\by_n^{\miss}$ and $\btheta$ is
\beq
O(NC^3).
\label{eq:order_theta}
\eeq

The most complex piece of the MCMC is updating the $\cB_{j}$ and $\cC_{j}$.  Each update of $\cB_j$ requires the construction of $\bm^*$ and $\bV^*$ in (\ref{eq:beta_update}) and then the generation of a multivariate normal deviate with mean $\bm^*$ and variance $\bV^*$.  The construction of $\bV^*$ requires multiplication of $\bPhi_j^T\bPhi_j$ and ${\bPhi_j^*}^T\bPhi_j^*$ which have computational complexity $O\{(L^\eta_j)^2N\}$ and $O\{(L^\eta_j)^2M\}$, respecitively.  There are also three $C \times C$ inverses and three Kronecker products of $C \times C$ matrices with $L^\eta_j \times L^\eta_j$ matrices, a ${(L^\eta_j} C)^2$ operation.  Thus the entire construction of $\bV^*$ has $O\{(L^\eta_j)^2(N+M) + ({L^\eta_j} C)^2\}$ complexity, and the same for $\bm^*$ by similar logic.  The dimension of $\bV^*$ is $L^\eta_jC$ and so the generation of a multivariate normal is $O\{(L^\eta_jC)^3\}$.  Hence the entire update of $\cB_j$ has complexity $O\{(L^\eta_j)^2(N+M) + ({L^\eta_j} C)^3\}$.  

By nearly identical arguments the complexity of a $\cC_k$ update is $O\{(L^\delta_k)^2N + ({L^\delta_k} C)^3\}$.  There are $J$ total updates of the $\cB_j$ and $K$ for the $\cC_k$.  Again, let $L=\max\{L^\eta_j,L^\delta_k\}$.  Then the complexity of the entire updating scheme for all $\cB_j$ and $\cC_k$ is 
\beq
O\{(J+K)(L^2(N+M) + (L C)^3)\}.
\label{eq:order_BC}
\eeq

Finally, putting the results of (\ref{eq:order_var}), (\ref{eq:order_theta}), and (\ref{eq:order_BC}) together, the complexity of an entire MCMC interation is 
$$
O\left\{MC^2 + NC^3 + (J+K)\left[LC^2 + L(N+M) + L^3C^3 \right]  \right\}.
$$
Ignoring the complexity induced by the number of output dimensions $C$ which is often small anyhow, the complexity can be written as
$$
O\left\{(J+K)\left[L(N+M) + L^3 \right]  \right\}.
$$
So while the complexity is now linear in $N+M$, the algorithm has gained complexity in the number of functional components for the emulator and discrepancy ($J$ and $K$) and the number of terms per functional component $L$.  While $L \approx 25$ is constant for most problems, $J$ and $K$ are not.  In particular, the $J$ depends on the sum of the dimensions of the input and parameter vectors $P+Q$.  Generally, $J=O\{(P+Q)^2\}$ for a two-way interaction model or $J=O\{(P+Q)^2+P^2Q\}$ for the limited three-way interaction model used in the results presented in the paper.  However, as long as the dimensionality of the model input and parameter spaces are moderate relative to the number of observations and simulator runs, this approach will have a large computational advantage over the existing approaches which are $O\{(N+M)^3\}$.

\vspace{-.15in}
\section{Multiple Models as Categorical Inputs}
\vspace{-.05in}
As mentioned previously, categorical parameters can be used to represent multiple competing models, and the uncertainty about which model is best for the situation.  For example, suppose there are two models, $\eta_1(\bx, \bt)$ and $\eta_2(\bx, \bt)$.  Conceptually there is just one simulator with an additional categorical parameter $t^*$, i.e., 
\vspace{-.15in}\beq
\eta(\bx, \bt, t^*) = 
\left\{
\begin{array}{ll}
\eta_1(\bx, \bt) & \mbox{if $t^*=1$} \\
\eta_2(\bx, \bt) & \mbox{if $t^*=2$} 
\end{array} 
\right.
\label{eq:mult_models}
\vspace{-.15in}\eeq
If it were assumed that $\eta_1$ and $\eta_2$ are somehow correlated GPs, then this construction would automatically borrow information from one simulator to help predict what another might output at an unevaluated input point. But this is exactly how categorical inputs are handled in the BSS-ANOVA model in Section~\ref{sec:cat_params}.

This framework can also be valuable in settings where there is a very computationally demanding simulator, but there is also a faster approximation (e.g., 2D or coarser grid, etc.) to complement the high-fidelity model.  In such situations the fast model(s) can be used to better approximate the more expensive model, provided it gives results somewhat similar to the expensive model.  Again, the categorical treatment automatically provides the correlation structure needed to leverage the information about the faster model to better predict the expensive model in locations it was not observed.  Also, as discussed previously the BSS-ANOVA framework provides the computational efficiency necessary to take advantage of many (e.g., 1,000 or more) runs of a fast, coarse model.

The representation of multiple models in (\ref{eq:mult_models}) only applies to a specific, yet common, case where the models have the same parameter sets.  In order to be completely general, a procedure to handle multiple models would have to allow for the parameter vectors $\bt$ to be different for different models.  The same framework could still be used, by grouping all distinct parameters into one large parameter vector. 
The main difficulty then becomes sampling between models, however.  The full development of a general multiple model approach is beyond the scope of this paper, but is a subject of further work.

\end{appendix}

\end{document}